\documentclass[final,3p,times,twocolumn]{elsarticle}

\usepackage{graphicx}
\usepackage{subfig} 
\usepackage[hyperindex,breaklinks]{hyperref}
\usepackage{breakurl}

\usepackage{amssymb}
\usepackage{amsthm}

\journal{Astroparticle Physics}

\begin{document}

\begin{frontmatter}

\title{Timing analysis techniques at large core distances for multi-TeV gamma ray astronomy}

\author{V. Stamatescu$^\dagger$}
\ead{victor.stamatescu@adelaide.edu.au}
\author{G. P. Rowell}
\author{J. Denman}
\author{R. W. Clay}
\author{B. R. Dawson}
\author{A. G. K. Smith}
\author{\\T. Sudholz}
\author{G. J. Thornton}
\author{N. Wild}

\address{School of Chemistry and Physics, The University of Adelaide, Adelaide 5005, Australia\\
         $\dagger$ now at Institut de F\'{i}sica d'Altes Energies, Edifici Cn., E-08193, Bellaterra (Barcelona), Spain}

\begin{abstract}
We present an analysis technique that uses the timing information of Cherenkov images from extensive air showers (EAS).
Our emphasis is on distant, or large core distance $\gamma$-ray induced showers at multi-TeV energies.
Specifically, combining pixel timing information with an improved direction reconstruction algorithm, 
leads to improvements in angular and core resolution as large as $\sim40\%$ and $\sim30\%$, respectively,
when compared with the same algorithm without the use of timing. Above 10 TeV, this results
in an angular resolution approaching $0.05^{\circ}$, together with a core resolution better than $\thicksim15$ m.
The off-axis post-cut $\gamma$-ray acceptance is energy dependent and its full width at half maximum
ranges from $4^{\circ}$ to $8^{\circ}$. For shower directions that are up to $\thicksim6^{\circ}$ off-axis,
the angular resolution achieved by using timing information is comparable, around 100 TeV, to the on-axis angular resolution.
The telescope specifications and layout we describe here are geared towards energies above 10 TeV.
However, the methods can in principle be applied to other energies, given suitable telescope parameters.
The 5-telescope cell investigated in this study could initially pave the way for a larger array of sparsely spaced telescopes
in an effort to push the collection area to $>10$ km$^{\mathrm{2}}$.
These results highlight the potential of a `sparse array' approach in effectively opening up the energy range above 10 TeV.
\end{abstract}


\end{frontmatter}

\section{Introduction}
\label{section0}

Based on current experimental evidence obtained from instruments such as H.E.S.S., MAGIC, CANGAROO-III, VERITAS and MILAGRO,
it is generally accepted that particles can be accelerated in shell-type supernova remnants to energies beyond a few $\times 10^{14}$ eV.
The H.E.S.S. instrument has observed a number of $\gamma$-ray sources with strong fluxes above 10 TeV \cite{Aharonian:2005,Aharonian:2006,AharonianRev:2008},
with some sources showing no evidence of a spectral cut-off.
Furthermore, most of the new sources found on the galactic plane have hard differential energy spectra with $\Gamma < 2.5$ \cite{Aharonian:2006}.
Future observations of $\thicksim100$ TeV $\gamma$-ray emission from these acceleration regions
will conclusively test current acceleration models and determine if SNR and/or other types of sources are responsible for cosmic ray acceleration up the \emph{knee}
\cite{AharonianNextGen:2005,Rowell:2005,AharonianRev:2008}.

An important advantage of the multi-TeV regime ($>$10 TeV) is that it becomes easier to distinguish between the hadronic ($\gamma$-rays from
cosmic-ray interactions) and leptonic ($\gamma$-rays from the inverse-Compton or IC process on soft photon fields) emission channels.
Electrons suffer strong energy losses due to synchrotron cooling if strong magnetic fields are present in their acceleration region.
At the same time, the Klein-Nishina effect reduces the IC cross section.
These effects strongly suppress the IC process so,
unless the $\gamma$-ray emission is associated with continuous or sporadic sources of multi-TeV electrons
(i.e. pulsar wind nebulae, active galactic nuclei),
observed spectra from steady and extended galactic sources that reach into the tens of TeV may favour a hadronic scenario.

The current generation of  Imaging Atmospheric Cherenkov Telescope (IACT) arrays are
limited by their collecting area in the multi-TeV energy range \cite{AharonianNextGen:2005,Rowell:2005}.
Good spectral measurements at energies approaching 100 TeV and beyond, assuming reasonable observation times ($\thicksim$ 50 h),
may require a dedicated instrument with collecting areas of around 10 km$^2$ \cite{AharonianNextGen:2005}.
Based on experience with  H.E.S.S, a wide field of view is also desirable,
both in terms of its survey potential and for the purpose of selecting off-source regions for cosmic-ray background determination in imaging large scale sources.
Meeting these design goals while maintaining a good angular resolution is important to the astrophysics potential of any such future instrument.
This is a key performance parameter that directly impacts on the point-source sensitivity and on the ability to probe the morphology of extended sources.

In this paper we investigate the use of shower timing information to improve stereoscopic shower reconstruction.
This follows from our earlier work \cite{Stamatescu:2008} and is motivated by our `sparse array' design \cite{Rowell:2008}, with which showers are often viewed
at large core distances (typically $>200$ m) by a cell of five telescopes, and, also by the first simulations of a multi-TeV cell by \cite{Plyasheshnikov:2000}.
The term `core distance' refers to the distance between the telescope and the shower axis, calculated in the shower plane.
In an alternative approach, Colin et al. \cite{Colin:2009} considered the performance of a dense IACT array over the energy range of 1 to 100 TeV.
Large-scale telescope arrays of varying densities, aimed at covering an energy range up to $\thicksim$100 TeV
are also under consideration for the planned CTA observatory \cite{cta_summary}.

The fluorescence technique \cite{Baltrusaitis:1985} pioneered air shower reconstruction methods of detectors such as Fly's Eye and HiRes,
in which angular and temporal information from a single shower image was combined to reconstruct the shower geometry.
For the atmospheric Cherenkov imaging technique,
the longitudinal and transverse temporal structures of EAS were first investigated using the HEGRA IACT system \cite{Hess:1999}.
By considering the image pixels times, it was found that a so-called \emph{time gradient} exists along the image major axis,
and that it is correlated with the core distance.
At small core distances ($\lesssim100$ m) this correlation is not present for hadronic showers,
and this difference, as suggested in \cite{Holder:2005}, is used by MAGIC \cite{Aliu:2009,Cabras:2008,Tescaro:2007} to improve background rejection.
This approach was also investigated using simulations of a wide field of view ($10^{\circ}$) camera \cite{Calle:2006},
using a single telescope as well as a 2-telescope system.
In that paper it was shown that the time gradient could be used to improve background rejection by a factor of $2-3$ for the case of a single telescope.
Although we do not attempted to link angular resolution directly with background rejection in this study,
we note that incorporating our timing analysis in a more comprehensive multi-parameter analysis,
such as that used by \cite{Calle:2006}, merits further investigation.

The \emph{time gradient} is also used by VERITAS to define an optimum signal integration window \cite{Holder:2005},
which improves the Cherenkov signal to skynoise ratio of the recorded signals.
Timing information was used for time-based image cleaning by the CANGAROO collaboration
(e.g. \cite{Kifune:1995,Kubo:2003}) in order reduce the effect of night sky background.
It is also used by MAGIC as a way of lowering the standard image cleaning thresholds \cite{Aliu:2009,Cabras:2008,Tescaro:2007}.

\section{Simulations of a 5-telescope cell for multi-TeV energies}
\label{section1}

Our simulations of EAS use the CORSIKA v6.204 \cite{Heck:1998} and SYBYLL \cite{Fletcher:1994} packages,
coupled with a telescope response simulation based on \emph{sim\_telarray} \cite{sim_telarray_paper,sim_telarray}.
Gamma-ray showers are generated over the energy range $1-500$ TeV using a flat $dN/dE \propto$ E$^0$ energy spectrum to enable good statistics
at high energies, and with a zenith angle of $30^{\circ}$. The chosen observation altitude of 220 m a.s.l. is typical of many Australian sites.

\begin{figure}
  \begin{center}
    \includegraphics[width=0.3\textwidth]{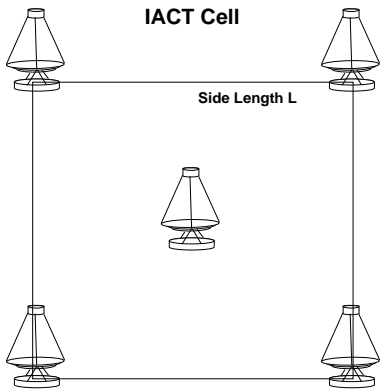}
  \end{center}
  \caption{
    Simulated cell layout: four telescopes are at the corners of a cell of side length $L=500$ m with the fifth telescope at the centre,
  }
  \label{fig:iact_cell}
\end{figure}

The simulated cell layout, shown in Figure \ref{fig:iact_cell}, is such that four telescopes are situated in the corners of a square of side 500 m,
while the fifth telescope is at its centre. Each simulated shower is used 20 times by scattering the core location
over a 1 km radius (in the shower plane) with respect to the centre of the IACT cell.
Simulations of the cell performance in which showers were re-scattered 10 times gave consistent results, in terms of
angular resolution as well as the mean and \emph{r.m.s.} of \emph{width} and \emph{length}, when compared to our default (20 times re-scatter) method.
We note the use of 10 times re-scattering of CORSIKA showers was successfully employed in the analysis of real H.E.S.S. data \cite{AharonianCrabNebula:2006}.
The collection area achieved with this cell layout approaches 1 km$^2$ above 10 TeV, and exceeds 1km$^2$ above 100 TeV \cite{Denman:2008}.

Each simulated telescope has $f/1.5$ optics with a modest-sized 6 m diameter
mirror of elliptic profile \cite{Schliesser}, of area 23.8\,m$^2$ (comprising $84\times60$ cm spherical mirror facets).
The circular camera has an $8.2^{\circ}$ diameter field of view (FoV)
made from 804 square pixels of side length $0.24^{\circ}$.
Previous simulation studies \cite{Aharonian:1995} found that a pixel diameter of $\thicksim0.25^{\circ}$
allowed for a good quality determination of the shower image second order moments (see discussion below),
while optimum image orientation parameters were achieved using $\thicksim0.15^{\circ}$ diameter pixels.
In the present study, the choice of pixel size is motivated by the size of the optical point spread function of each telescope.
Ray-tracing results \cite{Stamatescu:2010} indicate that the elliptic dish shape provides
a well-centred off-axis blur spot and  $\lesssim 0.25^{\circ}$ focussing out to $\thicksim 4^{\circ}$ off-axis.
Compared to a traditional Davies-Cotton design, this dish shape gives a small improvement of $\thicksim10$\%
in the $4^{\circ} - 5^{\circ}$ off-axis angle range.

A MODTRAN \cite{MODTRAN} tropical atmospheric profile and maritime haze model are used to treat the wavelength-dependent attenuation of Cherenkov light by the atmosphere.
In addition to the atmospheric transmission, \emph{sim\_telarray} treats the detection of photons by modelling the mirror reflectivity,
the photomultiplier (PMT) collection efficiency and the photo-cathode quantum efficiency, as discussed in \cite{sim_telarray_paper}.
The resulting photoelectron (\emph{pe}) signal is then input into
a detailed electronics simulation that models the data acquisition and telescope trigger.

Single \emph{pe} pulses (after the PMT, pre-amplifier and shaping amplifier)
are added in time over a 100 ns buffer such that the median Cherenkov \emph{pe} arrival time
is $\thicksim40$ ns after the start of the memory buffer.
The resulting signal in each data acquisition channel, which is AC-coupled and includes fluctuations from night sky background (NSB)
and electronic noise, is continuously sampled at 1 GHz and digitized using a 12 bit ADC (analog to digital converter).
An analog memory buffer would allow for this data acquisition scheme to be implemented in hardware.
A low gain channel is used if, at any given time, the digitized signal of the high gain channel exceeds the dynamic range of the ADC.

The pixel signal values and their corresponding pulse arrival times are obtained using a sliding integration window.
The highest digitized value in the time buffer is found, the digitized signal in a 21 ns region around the peak is integrated
and the number of \emph{pe} in the channel is determined. This integration time is sufficiently long to capture a single PMT pulse
for the range of core distances relevant to this study. The time of any given pixel (with respect to the start of the time buffer)
is given by the time bin that corresponds to the digitized signal peak.

Single \emph{pe} fast pulses are added in the time domain to test if a channel triggers its discriminator.
These produce output signals that are subsequently summed to determine whether a camera has triggered. 
A trigger signal for each telescope results if the signal in 3 near-neighbour (grouping of 9) pixels exceeds 6 \emph{pe},
and the centre pixel in the grouping is compulsory.
This condition is based on simulations \cite{Dunbar} that indicated
a low accidental single telescope trigger rate ($\thicksim$10$^{-2}$ Hz) due to NSB,
and the need to accept as many events as possible in order to obtain a large effective area.
A software stereo trigger is required for an event to be reconstructed.

A standard two-level image cleaning method \cite{Punch} is used to reduce the effect of NSB on shower images.
Pixels with signal over the (higher) image (or picture) threshold are included in the image,
while those over the (lower) boundary threshold must have a next neighbour above the image threshold to be included.
A previous investigation of the effect of image cleaning on the cell performance \cite{Denman:2008}
suggested a choice of 8 \emph{pe} image and 4 \emph{pe} boundary thresholds.

Shower images are  parametrized using the Hillas formalism \cite{Hillas},
which yields the Hillas parameters: a major axis, an image centre of gravity (\emph{cog}),
and shape parameters \emph{width} and \emph{length}, which are second order moments
in the direction perpendicular and parallel to the major axis, respectively.
The parameters prescribe an image as an ellipse. The \emph{image size} parameter is the sum of the \emph{pe} in the pixels that pass cleaning.

Given that a shower image is a two dimensional projection of the shower, the major axis is an approximate projection of the shower axis in the camera plane.
Therefore, if the image is well parametrized, the major axis will pass close to the true shower direction in the camera plane.
The pointing direction of the major axis is not prescribed by the Hillas formalism.
This axis `pointing degeneracy' may be broken if another major axis from a second telescope is available,
since, for well parametrized images, the axes will intersect in the vicinity of the shower direction in a common camera plane \cite{stereo}.

Shower reconstruction is performed for events in which at least 2 telescopes trigger and pass the following quality cuts:
an \emph{image size} cut of $>60$ \emph{pe} and a $dis2$ cut of $<4^{\circ}$.
$dis2$ is the maximum modulus value the of the vector sum:
\begin{equation}\label{eq:dis2}
  dis2 =  max(|\vec{cog} \pm \vec{axis} \,length|)
\end{equation}
where $\vec{cog}$ is the angular distance from the camera centre to the \emph{cog}, and $\vec{axis}$ is the unit vector of the image major axis.
The cut on $dis2$ mitigates the camera edge effects on shower images, and further optimization of this this cut is a topic of a future study.

\section{Shower reconstruction}
\label{section2}

Using our standard event reconstruction, the shower direction is determined by a weighted mean of the geometric intersection points of the major axes
in a common camera frame. This scheme, which was developed for the HEGRA IACT system, is denoted Algorithm 1 in \cite{alg3}.
The weights used in calculating the mean of intersection points are given by $w$:
\begin{equation}\label{eq:weights}
  w = (\textrm{\emph{size}}_{a} + \textrm{\emph{size}}_{b})^{2} \, \sin(s)
\end{equation}
where $a$ and $b$ denote two images used to obtain the intersection point, and $s$ is the so-called stereo angle between the two major axes.
This gives more weighting to brighter image pairs, which will, as a result of better Hillas parameterization, have more accurate major axes. This choice also
gives less weight to major axis pairs that are nearly parallel, which results in poor direction reconstruction \cite{alg3}.
The weights $w$ do not influence the reconstruction if only one pair of major axes is available (i.e. $n_{tel} = 2$).
Once the reconstructed shower direction is found, the core position
is determined using a similar method, by intersecting major axes
in the reconstructed shower plane, starting from the positions of the telescopes \cite{stereo}.

\begin{figure} 
  \begin{center}
    \includegraphics[width=0.48\textwidth]{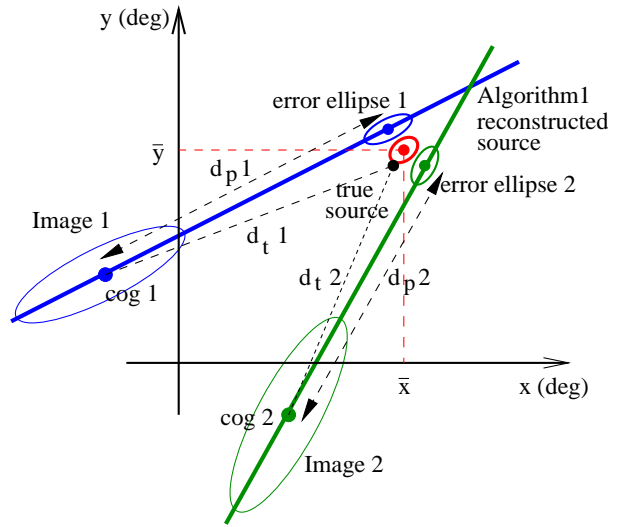}
  \end{center}
  \caption{The case of only two available images in an event reconstruction.
    The Algorithm 1 direction reconstruction is given by the intersection of the two major axes.
    Two error ellipses are combined using Algorithm 3 to give an improved estimate
    ($\mathbf{X}=(\overline{x},\overline{y})$) of the true source position.
    The predicted distance $d_{p}$ for each image is used to prescribe the angular displacement of each error ellipse from
    its respective \emph{cog}, along each major axis.
    The true distance, $d_{t}$, is the angular distance between the true source position and the image \emph{cog}.
  }
  \label{fig:alg3_combine}
\end{figure}

We also employ an improved method of reconstruction known as Algorithm 3 \cite{alg3},
which was also developed for HEGRA and which has been used in H.E.S.S. analysis (e.g. \cite{Aharonian:2007}).
Algorithm 3 is illustrated in Figure \ref{fig:alg3_combine} for the case of only one pair of major axes.
For each shower image, the expected uncertainties in the major axis orientation and the \emph{cog} position are used to
define an error ellipse. These uncertainties are obtained from lookup tables with dependencies on \emph{image size} and \emph{width/length}. 
The lookup tables are generated from simulations, using a flat $dN/dE \propto$ E$^0$  spectrum,
by considering the positional differences between the parameterized major axis of each image
and the true shower axis projected into the camera plane \cite{Stamatescu:2010}.
This gives an estimate of the true source position together with errors associated with the estimate.
The error ellipses are then combined using a weighted mean (e.g. see \cite{err_ellipse,Stamatescu:2010}),
where the weights are chosen to minimize the error ellipse associated with the Algorithm 3 result
($\mathbf{X}=(\overline{x},\overline{y})$ shown in Figure \ref{fig:alg3_combine}).
The errors used to prescribe each each error ellipse are then re-determined by using this improved source estimate
and the weighted mean is re-calculated. The Algorithm 3 source position usually converges to a stable point
after two or three iterations.

The quantity $d_{p}$, which is illustrated in Figure \ref{fig:alg3_combine}, is used to prescribe the position
of each error ellipse on its corresponding major axis. It is the expected angular distance between the image \emph{cog}
and the true source position, and is traditionally predicted using combinations of image parameters,
such as \emph{length} or \emph{width/length} coupled with \emph{image size}.
The accuracy and precision with which $d_{p}$ is predicted is a contributing factor to the performance of Algorithm 3.

\section{The Cherenkov image time gradient}
\label{section3}

\begin{figure}
  \begin{center}
    \includegraphics[width=0.48\textwidth]{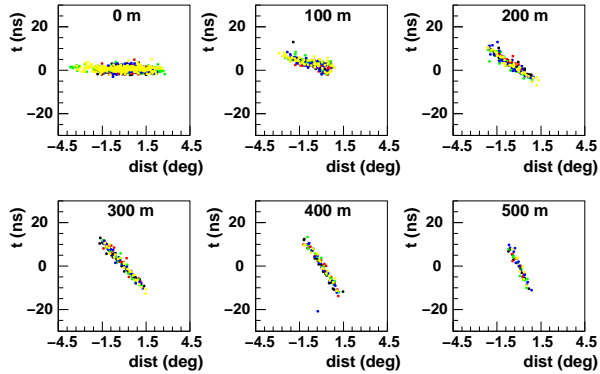}
  \end{center}
  \caption{Time-profiles of five 30 TeV $\gamma$-ray showers simulated from zenith and
    imaged at several core distances, which are indicated on each panel.
    Individual time-profiles are denoted by different colors.
    The pixel time \emph{t} is with respect to the weighted (using pixel amplitude) mean of image pixel times,
    whilst \emph{dist} is the projected distance along the major axis of each pixel with respect to the \emph{cog}.
  }
  \label{fig:tprofiles}
\end{figure}

The temporal and angular structure of an air shower
may be sampled by using the pixel times and their angular camera plane coordinates, respectively.
Figure \ref{fig:tprofiles} shows the time-profiles of five example 30 TeV $\gamma$-ray showers.
For each shower image, it shows the pixel time, $t$, versus $dist$,
the angular distance of each pixel from the \emph{cog}, projected along the major axis.
Here the $t$-axis origin corresponds to the pixel amplitude-weighted mean of recorded pixel times.
Figure \ref{fig:tprofiles} clearly shows that the average slope of individual time-profiles
tends to increase with a larger distance between the telescope and the shower core.

The EAS time development detected by IACTs was investigated by Hess et al. \cite{Hess:1999}.
It was found that when telescopes are located within the Cherenkov \emph{shoulder} ($\thicksim150$ m \cite{Hillas:1996}),
the detected photons that are emitted by ultra-relativistic secondary particles low in the atmosphere tend to arrive early in time compared 
to those emitted near the top of the shower. In our study the focus is on $\gamma$-ray showers detected at large core distances.
For showers detected beyond $\thicksim150$ m from a telescope, photons emitted near the top of the shower
tend to arrive before those emitted near the bottom.
We have investigated this effect for showers up to $10^{\circ}$ off-axis,
a limit which is imposed by our field of view.
Showers viewed at large core distances are also more extended in time, which can be seen in Figure \ref{fig:tprofiles}.
The \emph{time gradient} parameter is determined for each Cherenkov image as the average slope of its time-profile.
In our analysis this is calculated using a weighted linear fit, where each weight is the relative signal strength
(i.e. number of \emph{pe} relative to the \emph{image size}) of an image picture pixel (with $>$ 8 \emph{pe}).

We use a three dimensional toy model of $\gamma$-ray shower longitudinal time development
to examine the correlation between \emph{time gradient} and core distance,
and the dependence of \emph{time gradient} on the height of maximum Cherenkov light emission.
It can also be used to test the effect of off-axis shower geometries, which indicates
that the \emph{time gradient} remains approximately constant
for a shower axis that is inclined with respect to the telescope axis by angles of $\lesssim 10^{\circ}$ \cite{Stamatescu:2010}.
The model is best applied to showers with core distance greater than  about 150 m
and provides a check on the simulations in the large core regime ($>200$ m from a telescope).

\begin{figure*}
  \begin{center}
  \subfloat[]{\includegraphics[trim = 0.5mm 0.5mm 0.5mm 1mm, clip, width=0.47\textwidth]{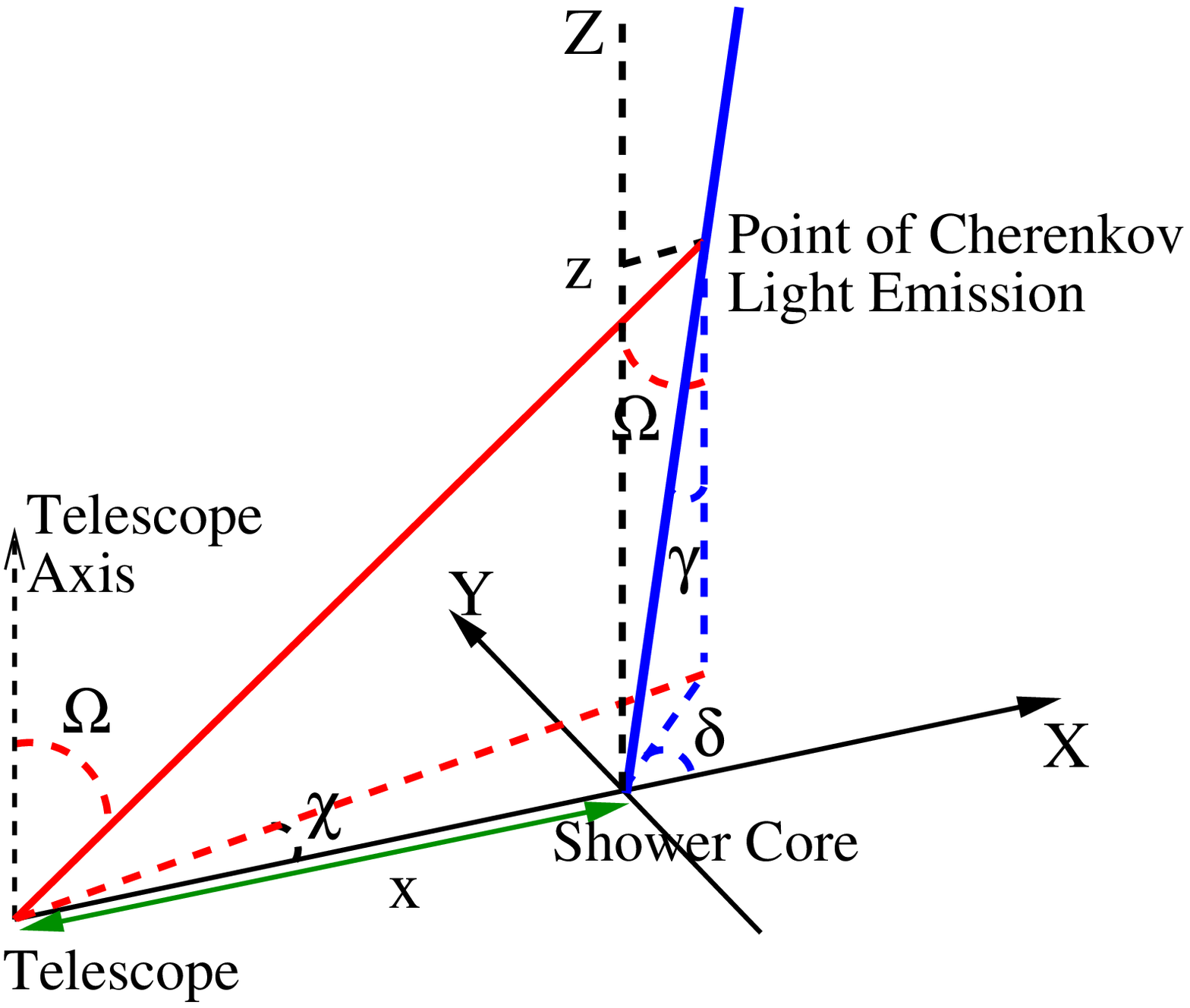}}
  \subfloat[]{\includegraphics[trim = 0.5mm 1mm 2mm 2mm, clip, width=0.32\textwidth]{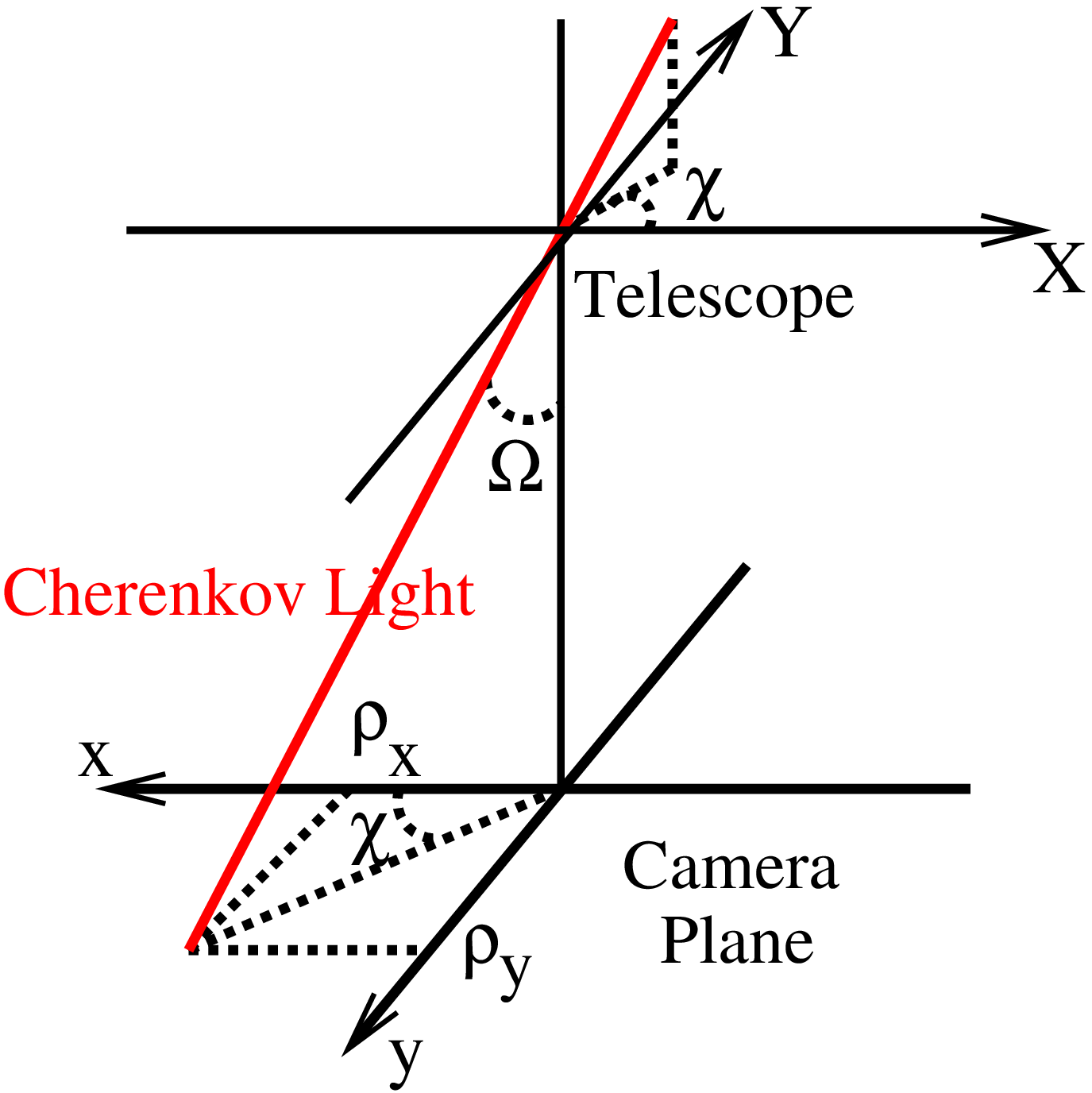}}
  \end{center}
  \caption{The geometry of time delay and imaging toy model:
    (a) A shower axis inclined by angle $\gamma$
    to the telescope axis and with azimuth angle $\delta$ lands at core distance $x$ from the telescope.
    Cherenkov light (shown in red) emitted at height $z$ on the shower axis, reaches the telescope inclined with respect to the vertically pointed
    telescope axis by $\Omega$ and with azimuth
    angle $\chi$.
    (b) Light whose direction is prescribed by the angles $\Omega$ and $\chi$, is imaged into the camera plane to a point ($\rho_x,\rho_y$).}
  \label{fig:toymodel}
\end{figure*}

The toy model handles the case of a vertically pointed telescope, illustrated in Figure \ref{fig:toymodel}.
Using an approach similar to that of Cabot et al. \cite{Cabot:1998},
we calculate the time delay $dt$ at the telescope, for a given core distance $x$:
\begin{equation}
  dt(x,z)=\frac{1}{c}\;\left(L - z/\cos(\gamma)\right)
  \label{eq:timeoff1}
\end{equation}
where $dt$ is the difference between the arrival time at the telescope of light emitted on the shower axis at height $z$
and the time at which the extrapolated primary $\gamma$-ray reaches the ground level, if it were propagating with speed $c$. 
The optical path $L$ in Eq. \ref{eq:timeoff1} is obtained by integrating the altitude-dependent
refractive index $\eta(h)$ along the path of the the Cherenkov light.
A simple dependence of the refractive index on altitude is assumed: $\eta(h) = 1+\eta_0\,e^{-h/h_0}$,
with $\eta_0=2.76\cdot10^{-4}$ and $h_0=8.0$ km (see \cite{Stamatescu:2010} for details).
Using angles $\gamma$ and $\delta$ as defined in Figure \ref{fig:toymodel} (a),
this leads to an expression for the time delay given by Eq. \ref{eq:timeoff} in \ref{appendix}.

Figure \ref{fig:toymodel} (b) shows that the light reaching the telescope is imaged into the camera plane
to a point with angular coordinates $ (\rho_x,\rho_y)$, given by:
\begin{equation}
  \rho_x(x,z)=-\Omega \,\cos\chi \qquad \rho_y(x,z)=-\Omega \,\sin\chi
  \label{eq:sigx}
\end{equation}
where the angles $\Omega$ and $\chi$ are determined by Eq. \ref{eq:omegaoff} and Eq. \ref{eq:chioff}, respectively in \ref{appendix}.

\begin{figure*}
  \begin{center}
  \includegraphics[trim = 8mm 9mm 16mm 22mm, clip, width=0.32\textwidth]{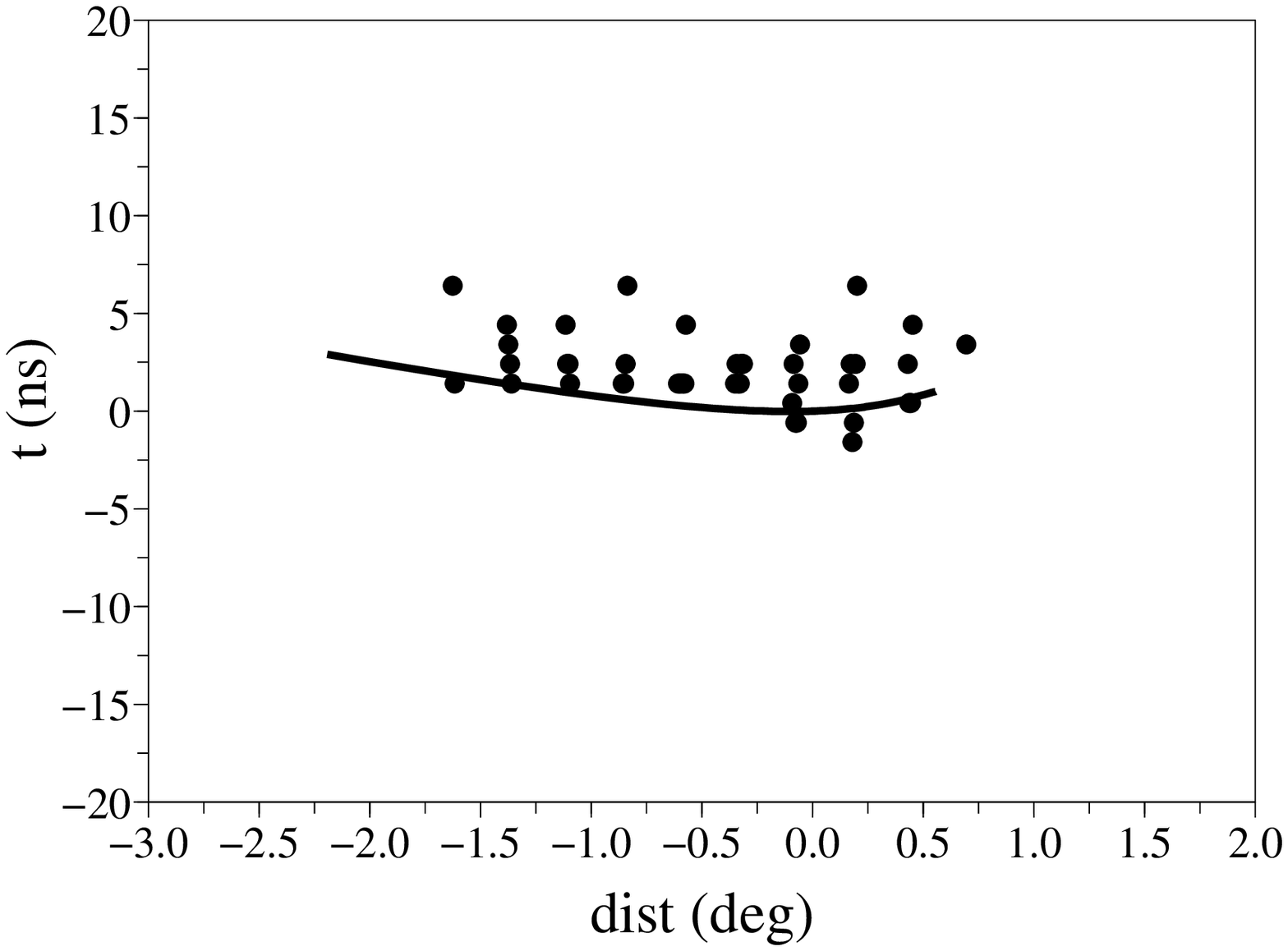}
  \includegraphics[trim = 8mm 9mm 16mm 22mm, clip, width=0.32\textwidth]{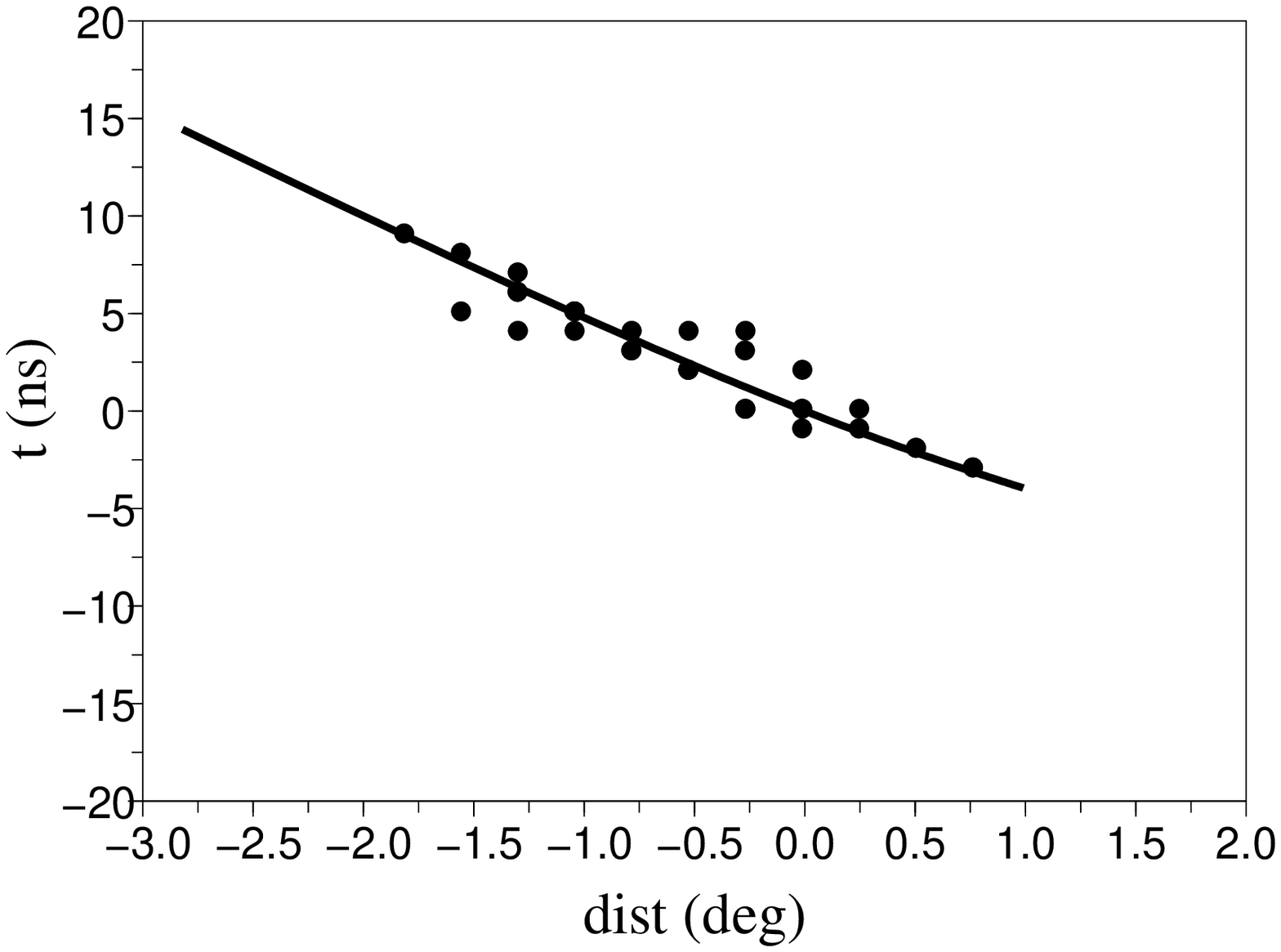}
  \includegraphics[trim = 8mm 9mm 16mm 22mm, clip, width=0.32\textwidth]{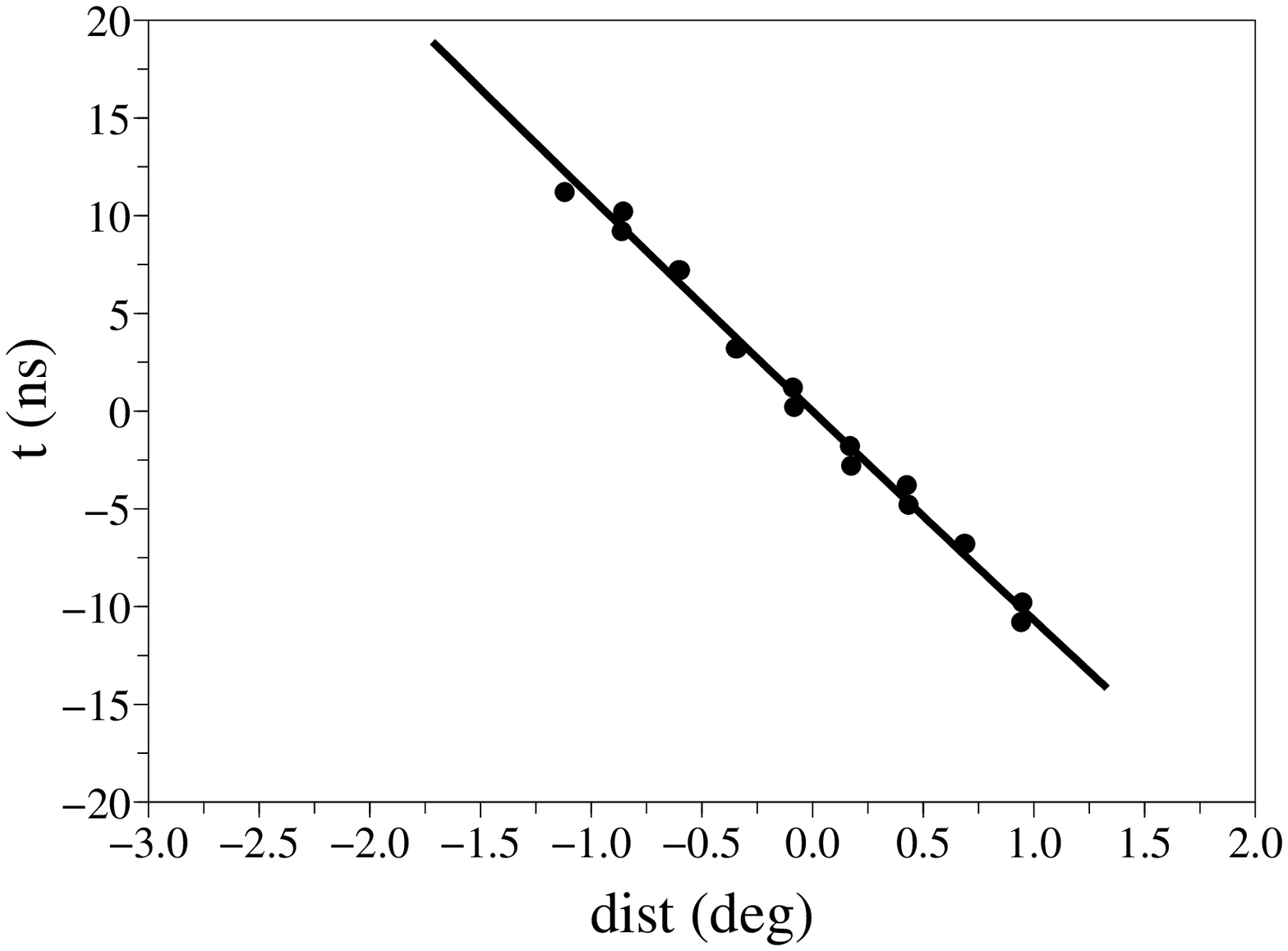}
  \end{center}
  \caption{Simulated pixel time $t$ (black points), with respect to the weighted mean time of image pixels, versus ($dist$),
    its distance along the major axis from the \emph{cog}, for an on-axis 30 TeV $\gamma$-ray shower from zenith
    with core distance from the telescope of 80 m (left), 200m (middle), 400m (right).
    Solid lines are the results of the toy model with angles $\gamma$ and $\delta$ set to $0$,
    and $z$ ranges from 1.5 km to 15.0 km (left), 2.5 km to 15.0 km (middle), 5.0 km to 15.0 km (right).
    The section of each line with the most positive values of $dist$ corresponds to the top of the shower axis.}
  \label{fig:tvsdist}
\end{figure*}

The results of the toy model are compared in Figure \ref{fig:tvsdist} with Monte Carlo simulated time profiles obtained for an on-axis 30 TeV $\gamma$-ray shower.
Differences between the model and the simulations become evident at smaller core distances ($<100$ m)
where the transverse development of the shower, which is not modelled, becomes important.
For the toy model, the origins of the $t$ and $dist$ axes correspond to the height of maximum light emission, $z_{max}$:
\begin{equation}
  \frac{1}{z_{max}}=\frac{\pi}{180}\,\frac{d_{t}}{x}
  \label{eq:rectruelightmax}
\end{equation}
where $d_{t}$ is the angular distance between the true source position in the camera and the image \emph{cog}, which is illustrated in Figure \ref{fig:alg3_combine}.
In choosing $z = z_{max}$ as the reference point, the $dist$-axis origin in Figure \ref{fig:tvsdist} corresponds, by definition, to the \emph{cog}.
For the $t$-axis, $dt(z_{max},x)$ is only approximately equal to the weighted mean time of image pixels
(w.r.t. the time of the extrapolated primary reaching the ground).
This assumption does not hold as well for showers with small core distances,
as seen from the the negative $t$-axis-offset of the curve with respect to the simulated points in Figure \ref{fig:tvsdist} (a).
However, the choice of axis origin does not alter the modelled time-profile shape, which is what determines the toy model \emph{time gradient}:
\begin{equation}
  \frac{\Delta dt}{\Delta \Omega} = \frac{dt(z+dz)-dt(z)}{\Omega(z+dz)-\Omega(z)}
    \label{eq:numdiff}
\end{equation}
We choose $dz = 0.1$ km, so that the quantity $\Delta dt/\Delta \Omega$ may be considered a \emph{local} \emph{time gradient},
which corresponds to a part of the shower around a given value of $z$.

\begin{figure}
  \includegraphics[width=0.48\textwidth]{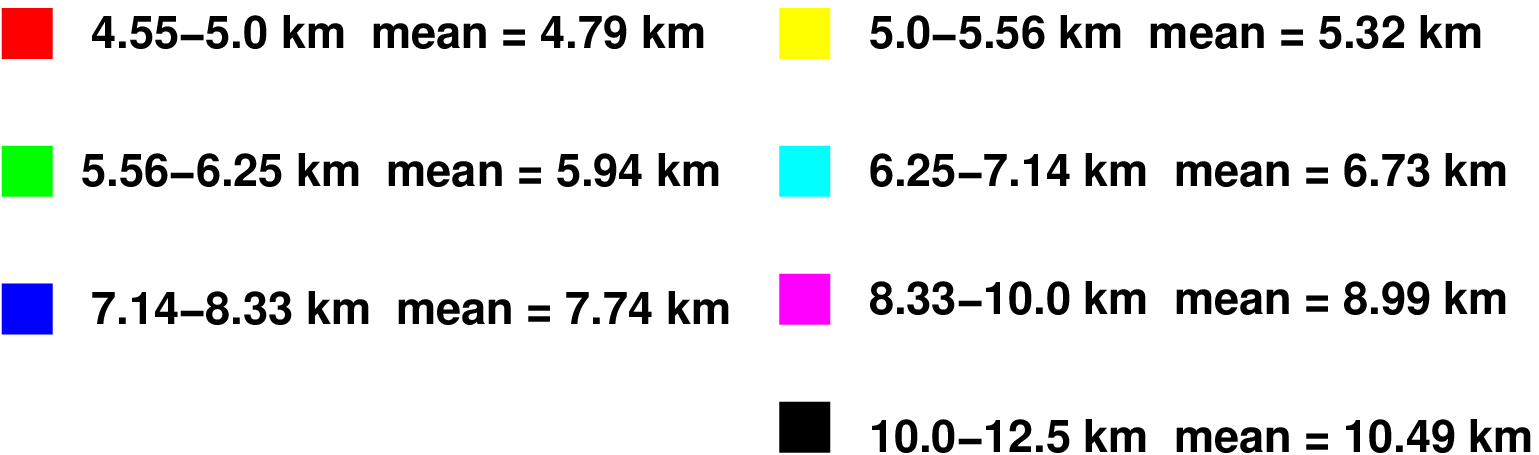}
  \includegraphics[trim = 5mm 100mm 10mm 12mm, clip, width=0.5\textwidth]{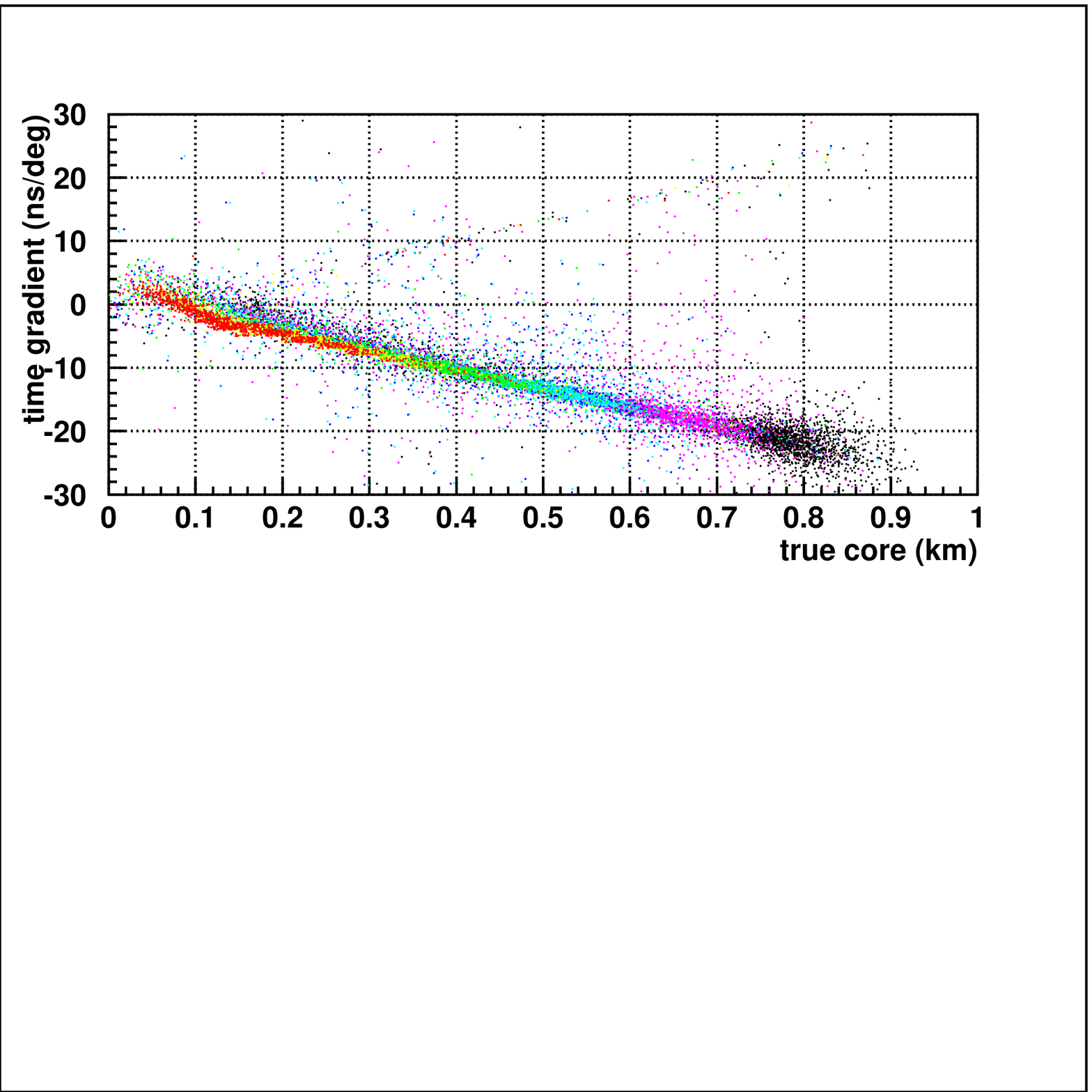}
  \caption{ The scatter plot of \emph{time gradient} versus true core distance, using the simulated $\gamma$-rays with primary energy between 1 TeV and 500 TeV.
    The major axis degeneracy is broken using Algorithm 1 (see text for details).
    The colors correspond to the indicated ranges of reconstructed light maximum, $h_{l}$.
  }
  \label{fig:tgrad_scatter}
\end{figure}

Hess et al. \cite{Hess:1999} found that the \emph{time gradient} depends strongly on the distance between the telescope and the shower core.
This is also apparent from  Figure \ref{fig:tgrad_scatter}, which shows the correlation between these quantities.
The `reflected' points in Figure \ref{fig:tgrad_scatter}, with positive \emph{time gradient} at core distances $\gtrsim200$ m,
are caused by events with poorly reconstructed Algorithm 1 source positions, which are used to break the pointing degeneracy
and result in the major axis pointing away from the true source position.
The data in Figure \ref{fig:tgrad_scatter} are color-coded according to height of maximum Cherenkov light emission,
which is calculated for each image as:
\begin{equation}
  h_{l} = \frac{r}{d_{r}\,\pi/180.}\,\cos(zen) - r\,\sin(zen)
  \label{eq:reclightmax}
\end{equation}
where $d_{r}$ is the angular distance in the camera between the \emph{cog} and Algorithm 1 reconstructed source position,
and where $r$ is the Algorithm 1 reconstructed core distance (in the reconstructed shower plane).

\begin{figure}
  \begin{center}
  \includegraphics[width=0.45\textwidth]{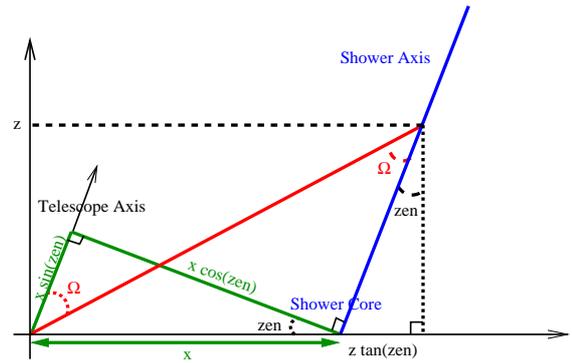}
  \end{center}
  \caption{Modified time delay and imaging toy model incorporating a zenith angle but simplified to handle only on-axis geometry.
    The x-axis is the line connecting the telescope to the shower core, with the telescope position taken as the origin.
    The \emph{true core distance} is taken as $x\;cos(zen)$, the distance between the telescope and the shower core projected onto the true
    shower plane. Light (in red) emitted at height $z$ from the shower axis reaches the telescope inclined with respect to the telescope axis by $\Omega$,
    which is measured as an angular distance in the camera.}
  \label{fig:toymodelzen}
\end{figure}

In order to compare $\Delta dt/\Delta \Omega$ with with our simulation results, we adjust the toy model to handle telescope pointing at non-zero zenith angles.
For simplicity, only on-axis shower geometries are considered and this is illustrated in Figure \ref{fig:toymodelzen}.
Time and angular development are modelled in terms of height $z$ and core distance $x$ by Eq. \ref{eq:timezen} and  Eq. \ref{eq:omegazen},
respectively in \ref{appendix}.

\begin{figure}
  \includegraphics[width=0.55\textwidth]{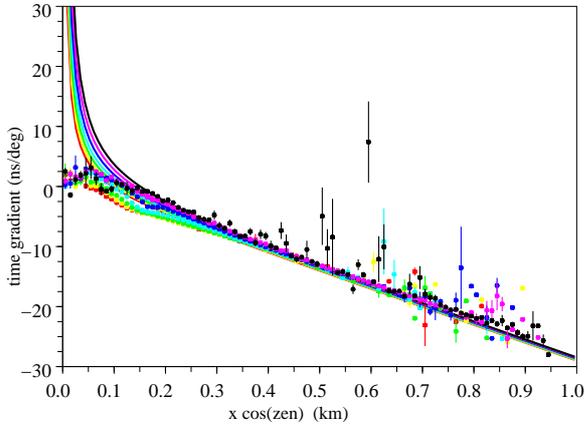}
  \caption{The \emph{time gradient} calculated from simulated data (points) and toy model (solid lines) are plotted against true core distance projected in the shower plane.
    The matching $z$ to $z + dz$ range in the model and $h_{l}$ range used to select data are color coded according to Figure \ref{fig:tgrad_scatter}.
    In the case of simulated data, the major axis pointing degeneracy, which affects the sign of the \emph{time gradient}, is broken using
    the true source position in the camera (this only applies to the data presented in this plot). 
  }
  \label{fig:tgrad_vs_core}
\end{figure}

The effect of $h_{l}$ on the \emph{time gradient} is modelled by choosing $z$ and $z+dz$ in Eq. \ref{eq:numdiff}
that match the corresponding range of $h_{l}$ used to split up the simulated data.
These ranges are given in Figure \ref{fig:tgrad_scatter} (b), together with a distribution of $h_{l}$ for our $\gamma$-ray data set.
Figure \ref{fig:tgrad_vs_core} shows that the toy model predictions ($\Delta dt/\Delta \Omega$)
and simulations are consistent for core distances beyond $\thicksim$200 m.
It also shows that in the large core distance regime the dependence of \emph{time gradient} on $h_{l}$ is weak.
The toy model \emph{time gradient} is not applicable for small values of $x$, where shower images become tend to be rounded and have poorly defined major axes.

\begin{figure}
  \begin{center}
    \includegraphics[trim = 5mm 100mm 100mm 12mm, clip, width=0.4\textwidth]{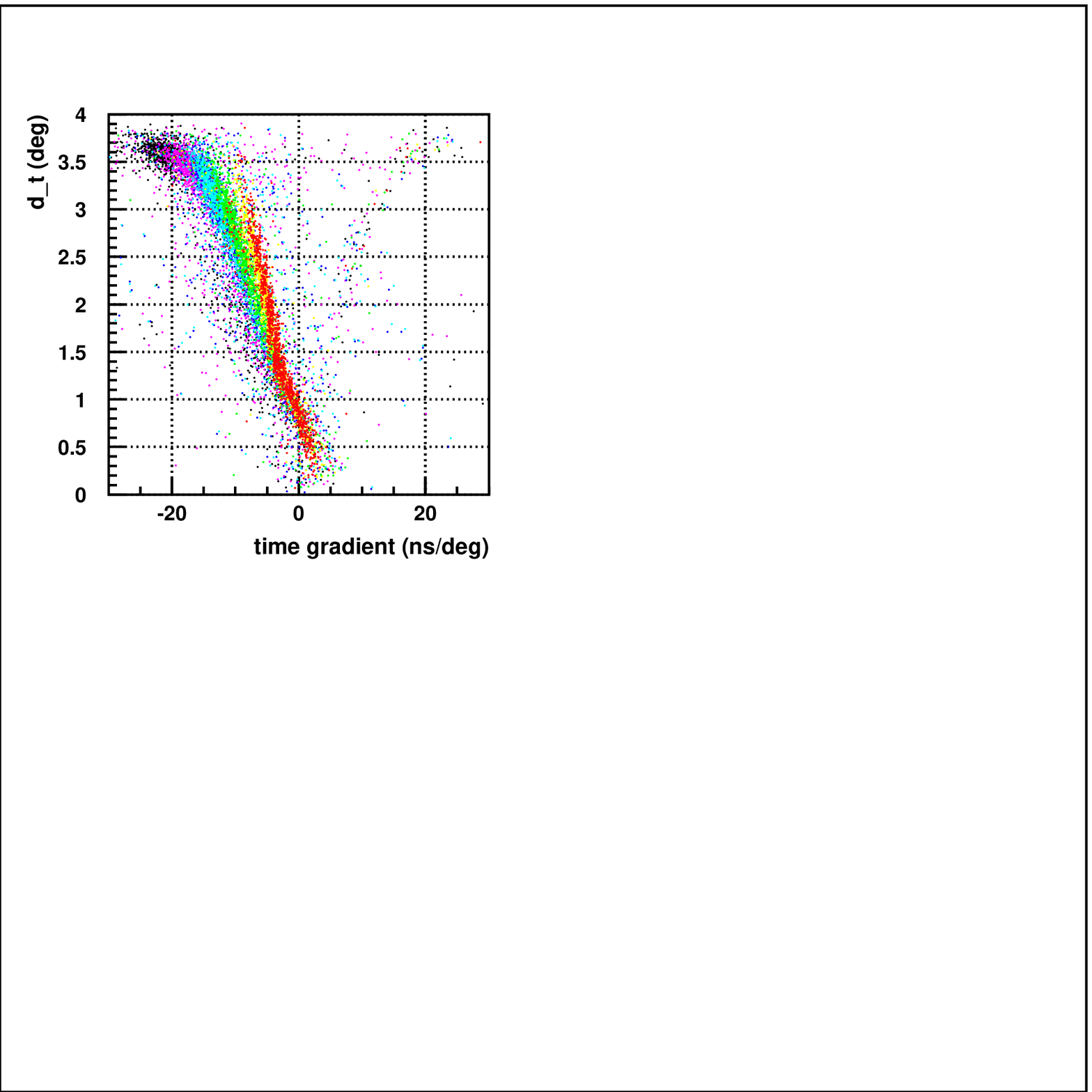}
  \end{center}
  \caption{The scatter plot of angular distance $d_{t}$ versus \emph{time gradient}, using the simulated $\gamma$-rays with primary energy between 1 TeV and 500 TeV.
    The colors correspond to the ranges of reconstructed light maximum, $h_{l}$, given in Figure \ref{fig:tgrad_scatter}.
  }
  \label{fig:tgrad_scatter2}
\end{figure}

Given that the angular distance $d_{t}$ will increase with core distance, a strong correlation also exists between $d_{t}$ and  the \emph{time gradient}.
This is evident in Figure \ref{fig:tgrad_scatter2} and suggests that the \emph{time gradient} may be used to predict the angular distance $d_{t}$.
From Figure \ref{fig:tgrad_scatter2} it is also apparent that the accuracy of this prediction can be improved if the
dependence of $d_{t}$ on the parameter $h_{l}$ is taken into account.
This is because a shower whose maximum emission is deeper in the atmosphere will result in a larger value of $d_{t}$.
Conversely a shower with the same core distance that has its maximum emission at a higher altitude will result in a smaller $d_{t}$.

A lookup table with dependencies on \emph{time gradient}
and  $1/h_{l}$ is filled with average values of  $d_{t}$.
This table is then used in the analysis to obtain the predicted angular distance $d_{p}$.
As detailed in Section \ref{section2}, this angular distance prescribes the position of a predicted source position along each major axis.
The predicted uncertainties, which are obtained using the error lookup tables, define an error ellipse around this predicted source position.
In this way, shower timing information is used in the Algorithm 3 direction reconstruction.

It worth noting that the optimal way in which to use pixel timing information may be
as part of a multi-parameter gamma-ray maximum likelihood reconstruction
(e.g. \cite{3Dmodelanalysis,modelanalysis}), rather than the piece-wise approach employed here.
However we leave this for future work.

\section{Cell performance using Algorithm 3}
\label{section4}

\subsection{Predicting the distance to the source}

\begin{figure*}
  \begin{center}
    \subfloat[]{\includegraphics[trim = 5mm 100mm 100mm 12mm, clip, width=0.34\textwidth]{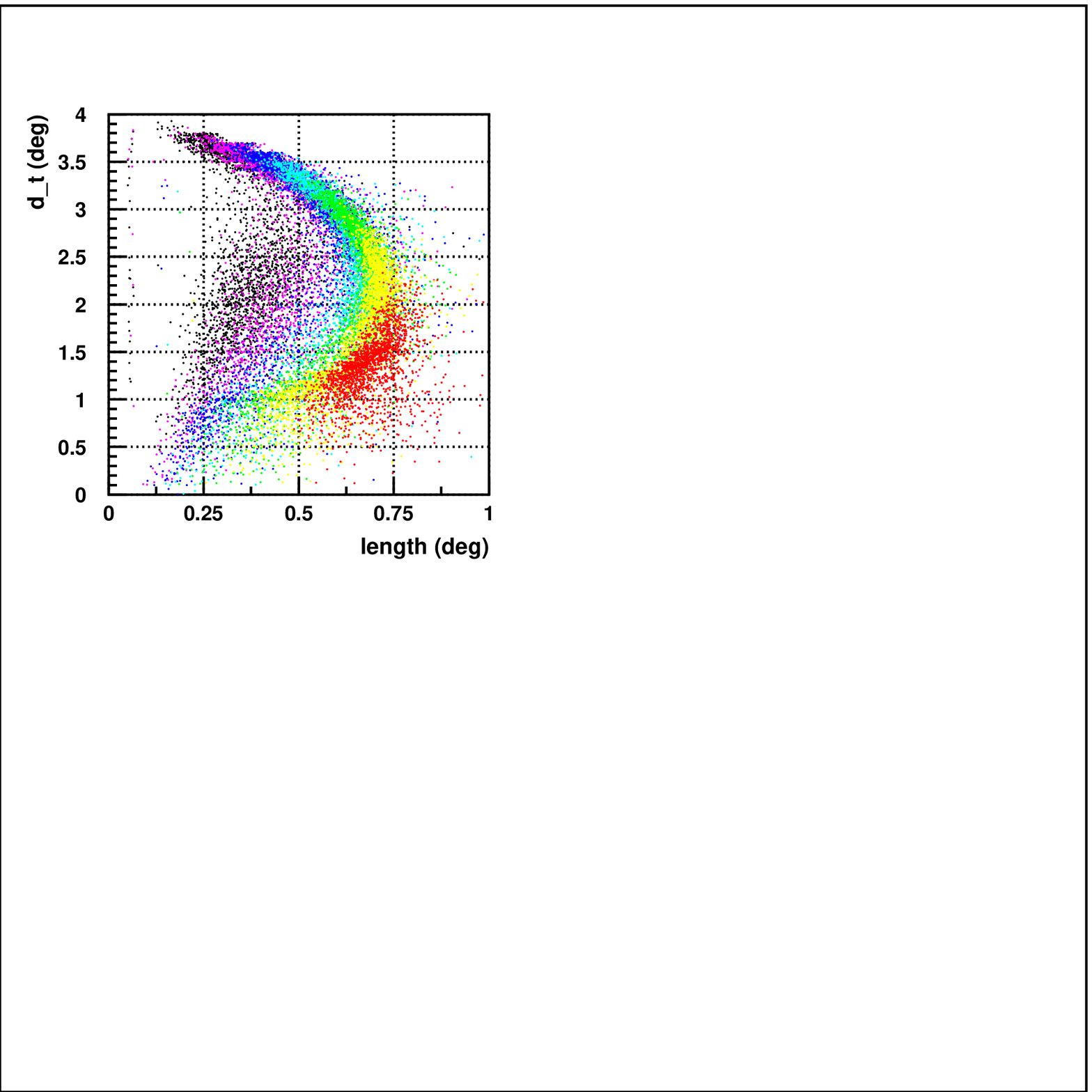}}
    \subfloat[]{\includegraphics[trim = 5mm 100mm 100mm 12mm, clip, width=0.34\textwidth]{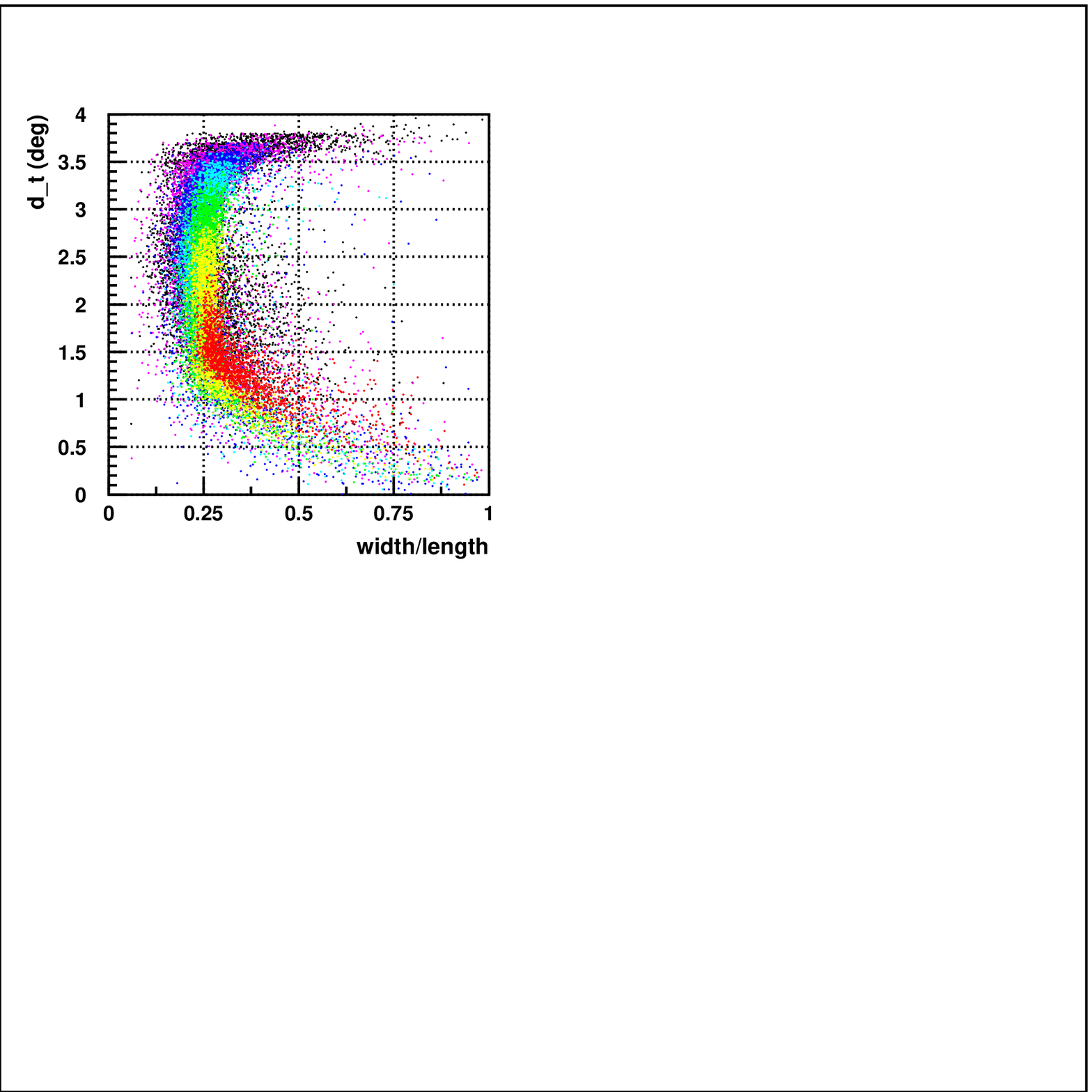}}
    \subfloat[]{\includegraphics[width=0.30\textwidth]{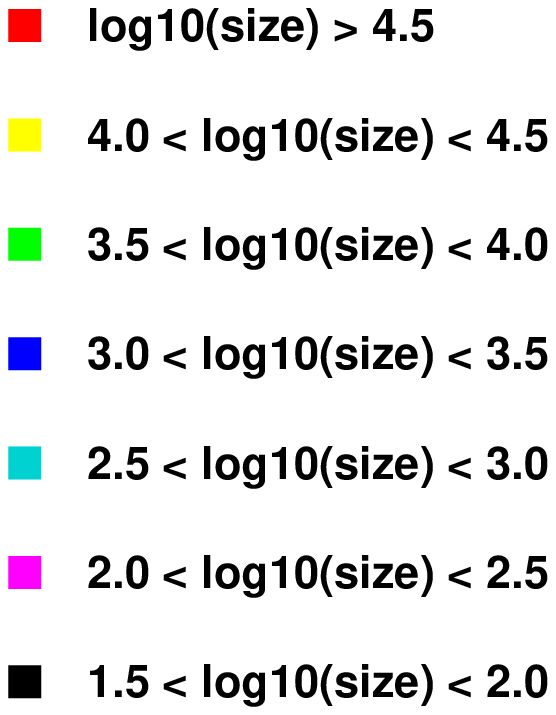}}
  \end{center}
  \caption{The true distance $d_{t}$ (deg.) plotted versus the predictors (a) \emph{length} and (b) \emph{width/length}.
    The colors denote ranges of log$_{10}$(\emph{image size}) given in (c).
  }
  \label{fig:dis_predictors}
\end{figure*}

In this study we compare the performance of Algorithm 3 using timing information with results obtained using traditional predictors of the distance $d_{t}$.  
The two lookup tables used for this comparison have dependencies on \emph{length} and log$_{10}$(\emph{image size})
and on \emph{width/length} and log$_{10}$(\emph{image size}), respectively.
Figure \ref{fig:dis_predictors} shows the behaviour of $d_{t}$ in terms of the dependencies of these lookup tables.

The \emph{length} of an image increases with core distance, which makes it a predictor of $d_{t}$.
A dependency on log$_{10}$(\emph{image size}) is included in this lookup table because the image \emph{length} increases with \emph{image size},
as shown in Figure \ref{fig:dis_predictors} (a).
The image aspect ratio, \emph{width/length},
decreases with increasing core distance as shown in Figure \ref{fig:dis_predictors} (b).
Both predictors are affected by camera edge effects for the distance range of $d_{t}$ $\gtrsim$  3$^{\circ}$,
which results in otherwise elongated images becoming more rounded.
This effect leads to smaller \emph{length} and larger \emph{width/length} values.

Both \emph{length} and \emph{width/length} suffer from a similar problem in predicting $d_{t}$,
which can be seen in Figures \ref{fig:dis_predictors} (a) and \ref{fig:dis_predictors} (b):
beyond  $d_{t}$ values of $\thicksim1.2^{\circ}$, each predictor is insensitive to changes in $d_{t}$.
Moreover, splitting the \emph{length}, \emph{width/length}, or \emph{length}$/$log$_{10}$(\emph{image size}) according to $h_l$
does not solve this problem due to the inter-dependent way in which $h_l$ and core distance influence the \emph{length}, \emph{width}, as
opposed to the \emph{time gradient} .

Beyond $d_{t}$ of $\thicksim2.75^{\circ}$ the predictors are well behaved,
but it is necessary to distinguish between this regime, and that of $d_{t}$ $\lesssim1.2^{\circ}$.
The \emph{width/length} table is therefore split into three separate lookup tables
that correspond to three ranges of $d_{t}$:  $<1.25^{\circ}$, between $1.25^{\circ}$ and $2.75^{\circ}$,
and $>2.75^{\circ}$. The \emph{length} lookup table is split in two ranges of $d_{t}$ $<2^{\circ}$ and $d_{t}>2^{\circ}$.
The choice of which lookup table to use for a given image is made based on the value $d_{r}$, which is initially obtained from the Algorithm 1 reconstruction.

Images with a small $d_{t}$ usually correspond to showers with small core distances,
whose time profiles tend to suffer from fluctuations \cite{Hess:1999}.
We therefore make an additional refinement when using the \emph{time gradient} and $h_{l}$ lookup table:
for images with $d_{r}<0.85$, the \emph{width/length} distance predictor is used instead.

\begin{figure*}
  \begin{center}
    \subfloat[]{\includegraphics[trim = 5mm 100mm 100mm 12mm, clip, width=0.32\textwidth]{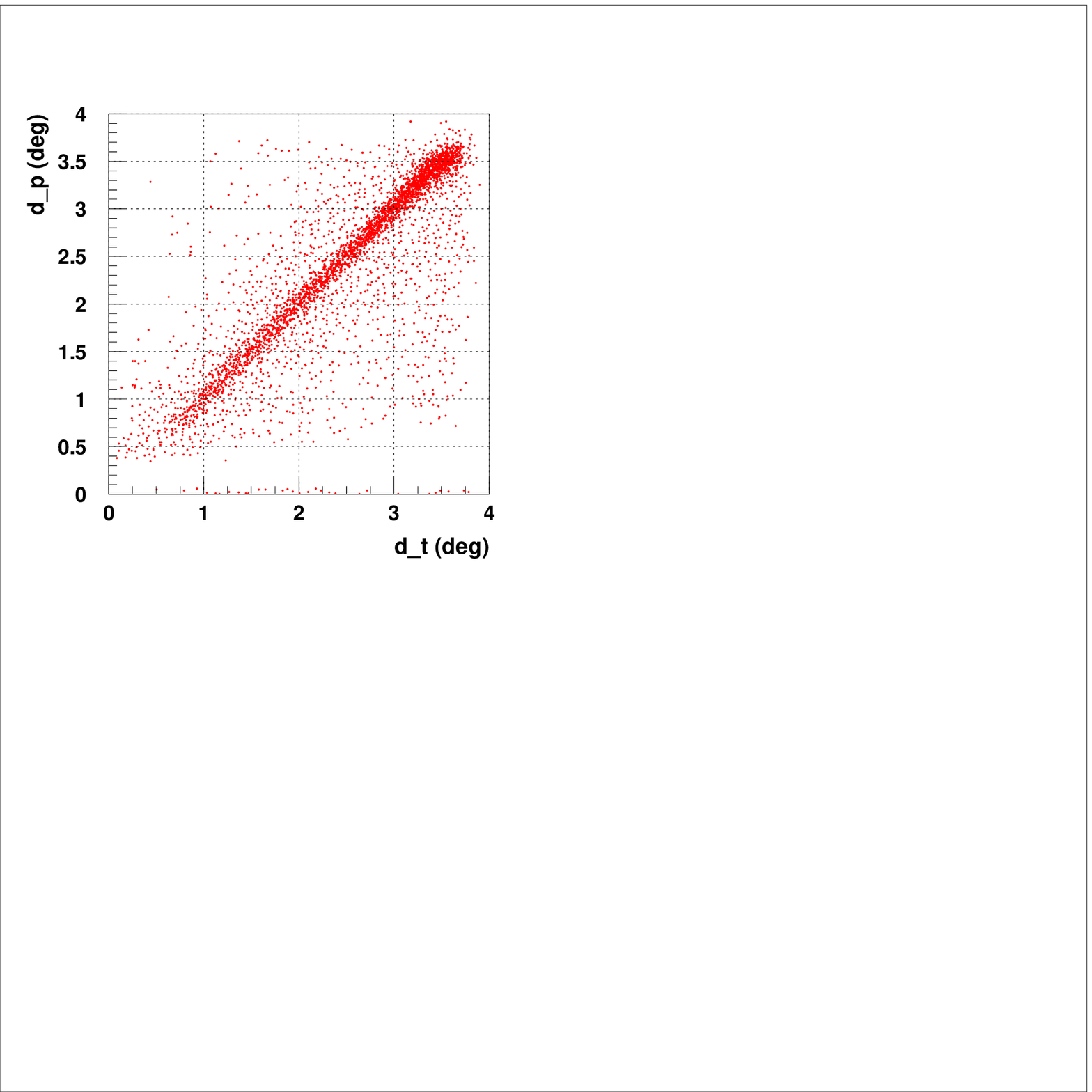}}
    \subfloat[]{\includegraphics[trim = 5mm 100mm 100mm 12mm, clip, width=0.32\textwidth]{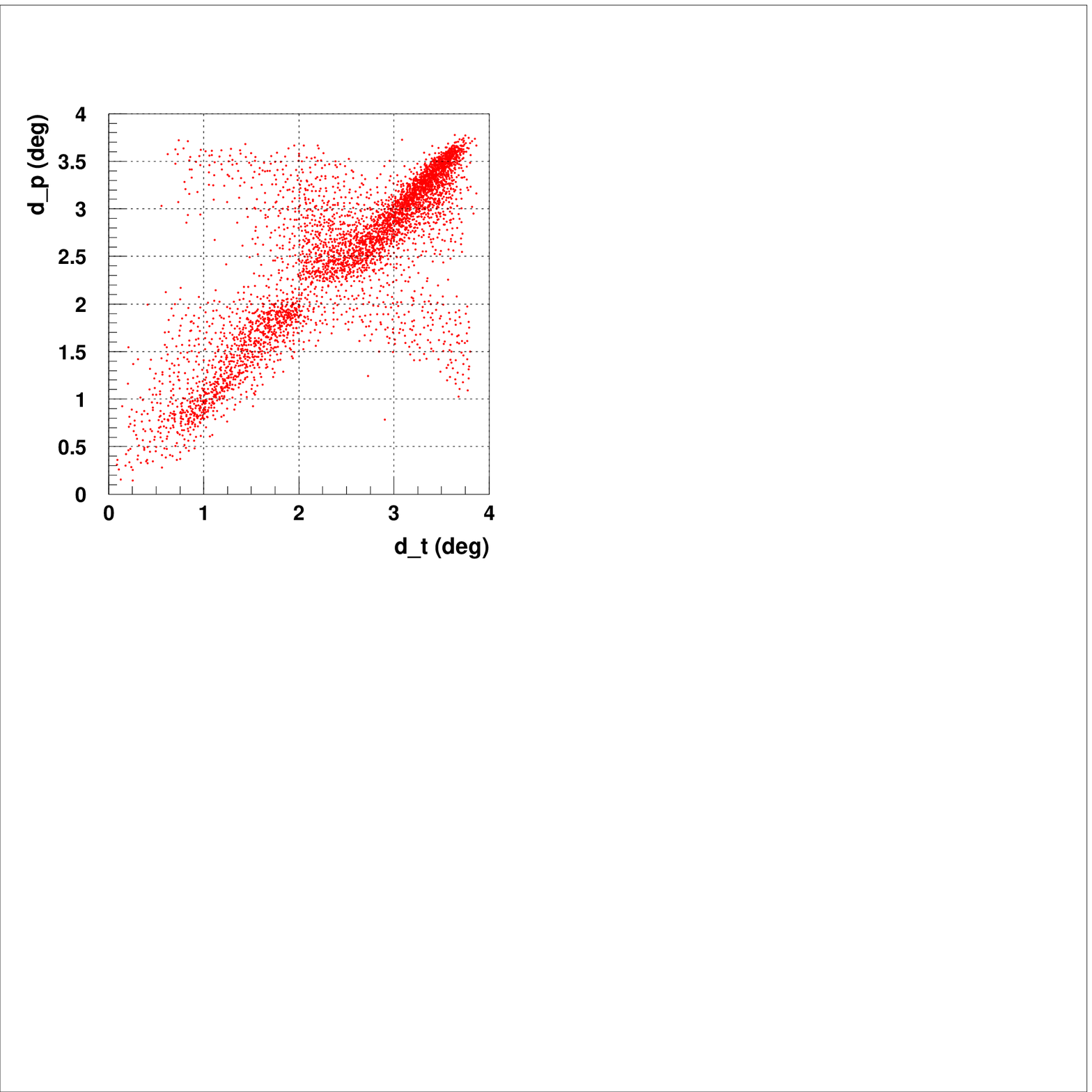}}
    \subfloat[]{\includegraphics[trim = 5mm 100mm 100mm 12mm, clip, width=0.32\textwidth]{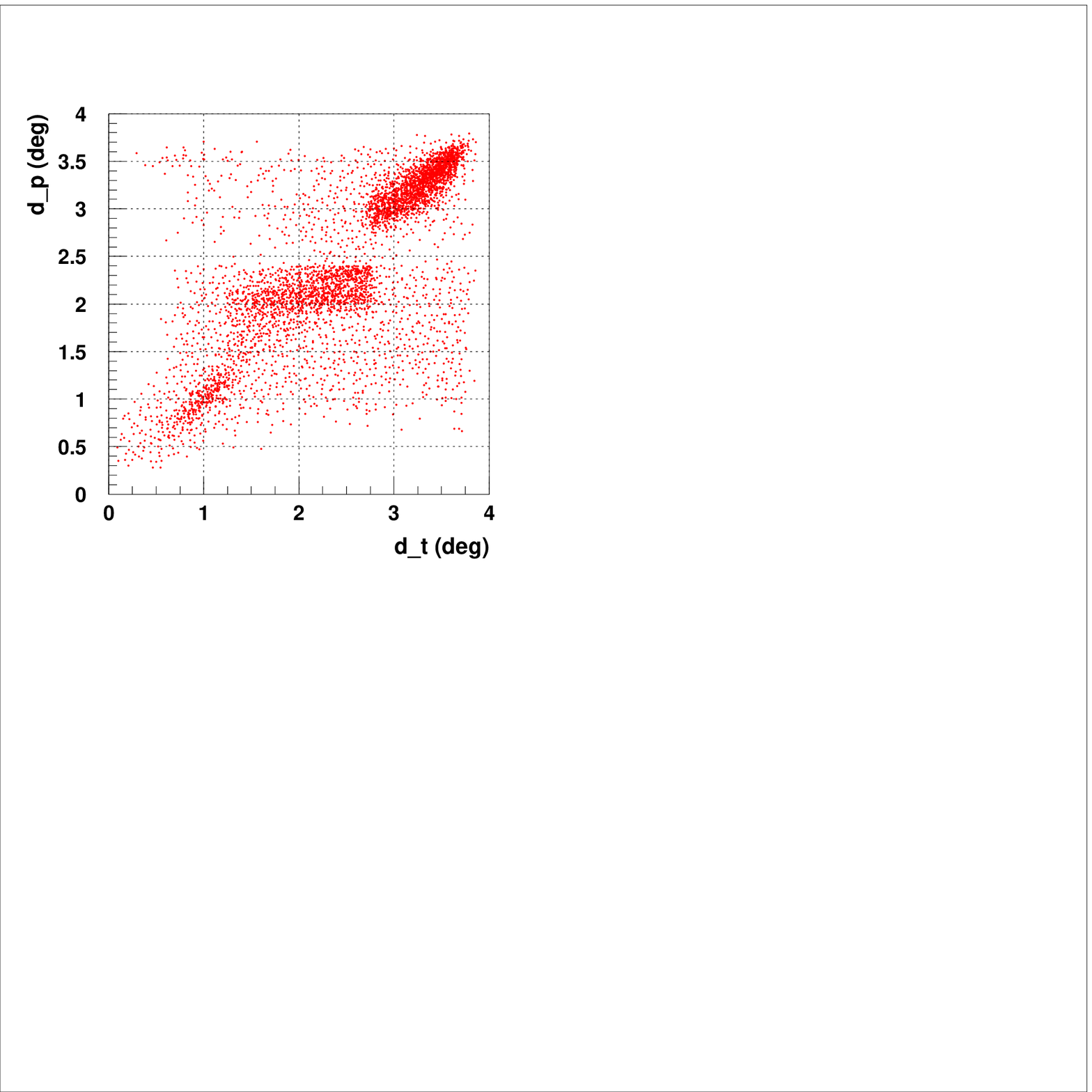}}
  \end{center}
  \caption{Predicted distance $d_{p}$ (deg.) plotted against true distance $d_{t}$ (deg.).
    $d_{p}$ is predicted using a distance lookup table 
    based on (a) \emph{time gradient} and 1/$h_{l}$, and \emph{width/length} for $d_{r}<0.85^{\circ}$,
    (b) \emph{length} and log$_{10}$(\emph{image size}), and (c) \emph{width/length} and log$_{10}$(\emph{image size}).
  }
  \label{fig:dis_predictors_results1}
\end{figure*}

The performance of the three sets of lookup tables dependencies is illustrated by Figure \ref{fig:dis_predictors_results1}.
The breaks in the scatter plots correspond to the aforementioned split ranges of $d_{t}$.
The combination of \emph{time gradient} and $h_{l}$ produces the strongest correlation (by eye) between
the true distance, $d_{t}$, and the predicted distance, $d_{p}$. 

\subsection{Angular resolution}

We evaluate the angular resolution of the telescope cell using $r68$, the angular radius containing 68\% of events.
We apply image shape cuts on mean-scaled parameters \cite{Konopelko_MSW}, which are used for cosmic ray background suppression.
The images that correspond to events passing the cuts are, on average, better parametrized, and this leads to an improved angular resolution.
The cut parameters are mean-scaled-width (MSW), mean-scaled-length (MSL) and mean-scaled-Npix (MSNPix) \cite{Rowell:2008}.
MSNPix is a new parameter, calculated by scaling the number of pixels in each image by the expected number of pixels
(obtained from simulations, for a given \emph{image size} and core distance)
and then averaging over the telescopes in the reconstruction.
The cut on MSNPix provides additional $\gamma$-hadron separation power (after MSW and MSL cuts) for energies above 50 TeV,
however it does not significantly affect the angular resolution.

\begin{figure*}  
  \subfloat[]{\includegraphics[trim = 5mm 125mm 100mm 12mm, clip, width=0.48\textwidth]{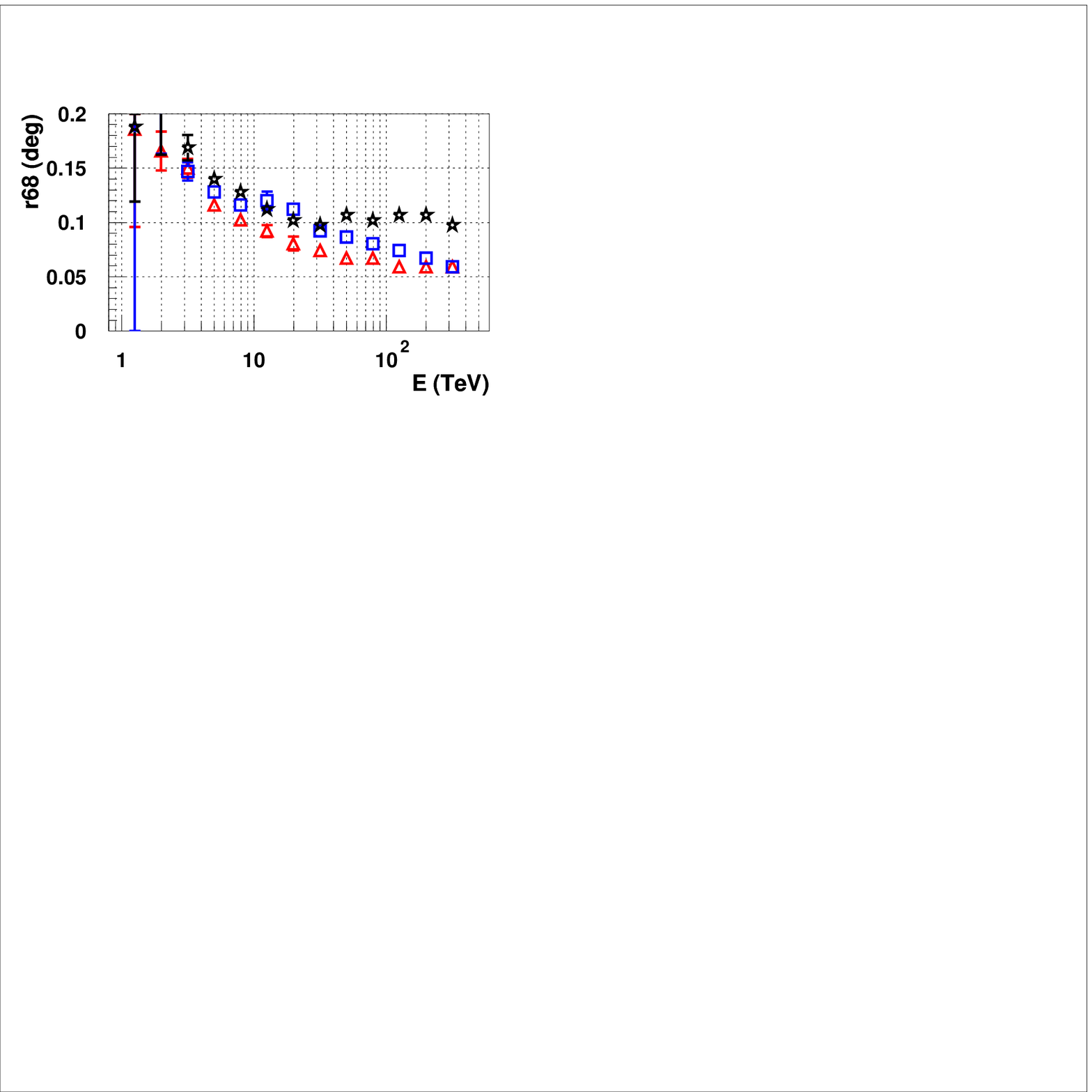}}
  \subfloat[]{\includegraphics[trim = 5mm 125mm 100mm 12mm, clip, width=0.48\textwidth]{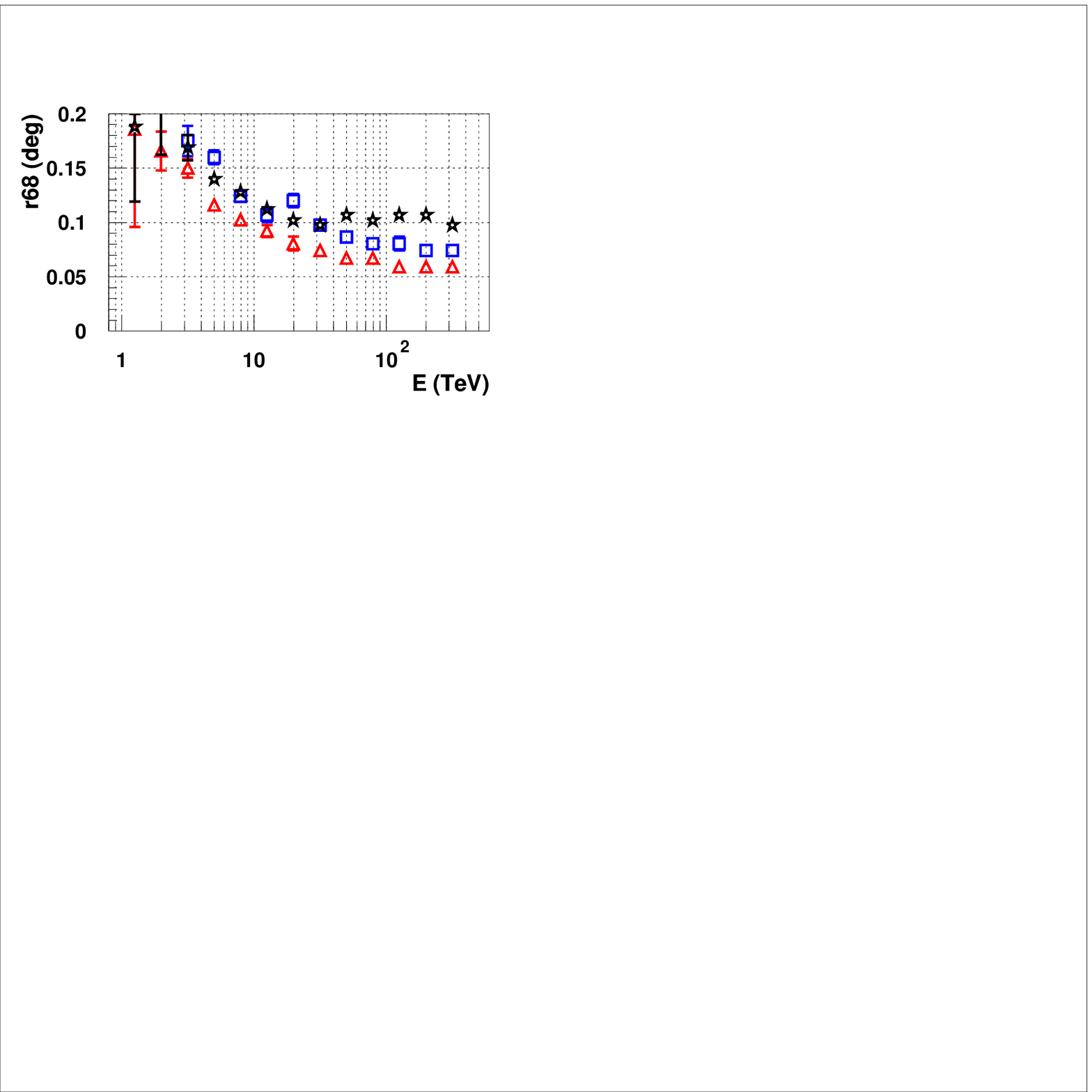}}
  \caption{Comparison of the angular resolution $r68$ (radius for 68 \% event containment) after all shape cuts (MSW$<1.05$, MSL$<1.2$, MSNPix$<1.1$),
    obtained using the \emph{time gradient} dependent Algorithm 3 implementation (red triangles),
    which uses predicted distance $d_{p}$ based on \emph{time gradient} and 1/$h_{l}$
    and two different versions of Algorithm 3 (blue squares) that use alternative methods for obtaining $d_{p}$: (a) $d_{p}$ is obtained using \emph{length} coupled 
    with log$_{10}$(\emph{image size}), and (b) $d_{p}$ is obtained using \emph{width/length} coupled with log$_{10}$(\emph{image size}).
    The Algorithm 1 results (black stars) are shown in both panels for comparison.
  }
  \label{fig:angres_super}
\end{figure*}

The post-cut (MSW$<1.05$, MSL$<1.2$, MSNPix$<1.1$) angular resolution obtained with the `\emph{time gradient} dependent' Algorithm 3,
which uses the \emph{time gradient} and $1/h_{l}$ lookup table,
is compared in Figure \ref{fig:angres_super} to that obtained obtained by using \emph{length}
or \emph{width/length} together with log$_{10}$(\emph{image size}).
For on-axis showers, the next-best results are achieved with the \emph{length} and log$_{10}$(\emph{image size}) lookup table,
and we term this the `\emph{length}-dependent' Algorithm 3.
The improvement in angular resolution given by the \emph{time gradient} dependent Algorithm 3 over the \emph{length}-dependent Algorithm 3 reconstruction
ranges from $\thicksim$10\% to $\thicksim$40\% in the energy range of $\thicksim$5 to $\thicksim$200 TeV.
The error bars are determined on the basis of Poisson fluctuations
in the number of events (68\% of events) used to calculate the radii for each energy bin.

Figure \ref{fig:angres_super} also shows, for comparison, the result of the purely geometric Algorithm 1 reconstruction.
The upturn in the Algorithm 1 angular resolution seen above $\thicksim$30 TeV is due to a worsening in the reconstruction
of low telescope multiplicity (2 or 3 telescope) events whose core distance from the array centre is large (greater than $\thicksim$300 m).
This subset of events suffers from smaller stereo angles (nearly parallel major axes), thereby increasing the error in major axis intersection points.
The Algorithm 3 direction reconstruction mitigates this problem.
For instance, the level of improvement provided by \emph{length}-dependent Algorithm 3 over the purely geometric reconstruction, 
is $\thicksim10\%$ below 100 TeV, and $\thicksim55\%$ above that energy.
Given its better distance prediction, the \emph{time gradient} dependent Algorithm 3 improves things further.
In this case, the level of improvement over Algorithm 1 is $\thicksim15\%$ below 10 TeV,
$\thicksim40\%$ in the 10 to 100 TeV energy range, and $\thicksim75\%$ above 100 TeV.

The simulation results indicate that by using timing information together with the Algorithm 3 direction reconstruction, 
an $r68$ angular resolution of less than $0.1^{\circ}$ is obtained above $\thicksim10$ TeV, and this value approaches $0.05^{\circ}$ at the highest energies.
This confirms the finding of an earlier study \cite{Calle:2006}, namely that a large collecting area can be achieved at multi-TeV energies
using a small number of telescopes, while maintaining good angular resolution.
Using our time-dependent stereoscopic analysis, we have demonstrated this for a `sparse array' of IACTs \cite{Rowell:2008}.
In the present study we show this to be case for off-axis performance (see the discussion at the end of this section).

\subsection{Core and energy reconstruction}

\begin{figure}
  \begin{center}
    \includegraphics[trim = 5mm 125mm 100mm 12mm, clip, width=0.48\textwidth]{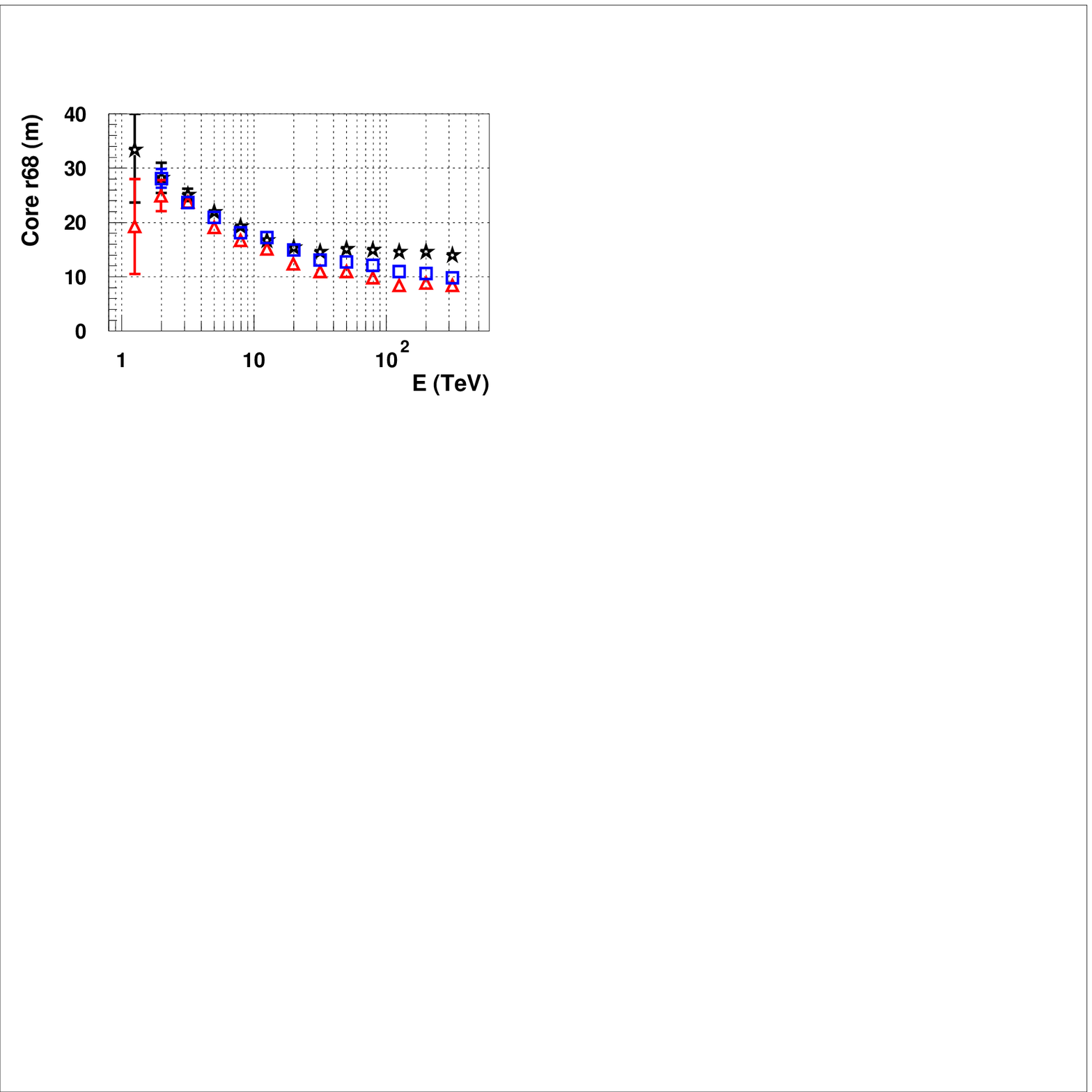}
  \end{center}
  \caption{Core Resolution r68 (radius for 68 \% event containment)
    as a function of true energy after shape cuts: MSW$<1.05$, MSL$<1.2$, MSNPix$<1.1$.
    The results shown as black stars use the Algorithm 1 core reconstruction.
    Those shown as red triangles use the \emph{time gradient} dependent Algorithm 3 core reconstruction,
    while those shown as blue squares use the \emph{length}-dependent Algorithm 3 core reconstruction.}
  \label{fig:coreres_alg3}
\end{figure}

The improved direction reconstruction provided by the \emph{time gradient} dependent Algorithm 3
can be used to improve the shower core reconstruction.
A new reconstructed core position is obtained by intersecting
axes defined using the \emph{time gradient} dependent Algorithm 3 reconstructed source position and the \emph{cog} of each image passing quality cuts.
These newly defined axes are intersected in the reconstructed shower plane, starting from their telescope positions.
This \emph{time gradient} dependent `Algorithm 3 core' resolution is compared in Figure \ref{fig:coreres_alg3}
to that obtained using the \emph{length}-dependent Algorithm 3 reconstruction.
With respect to the \emph{length}-dependent Algorithm 3 core resoltuion, an improvement of $\thicksim20\%$ is achieved
for energies above a few TeV through the introduction of timing information.
The purely geometric Algorithm 1 core reconstruction is also shown for comparison. 

Our shower energy reconstruction is performed using a $\gamma$-ray filled lookup table
with dependencies on \emph{image size} and the reconstructed core distance to each telescope.
The events used to fill the lookup table are first re-weighted in order to smooth out effects
of the large discontinuity in our Monte Carlo statistics at 10 TeV and 100 TeV.
An energy estimate is obtained for each telescope, and the estimates from telescopes used in the shower reconstruction
are combined to give a global energy estimate, $E_{reco}$.
This simple approach does not account for the effect of the height of shower maximum on the reconstructed energy,
that is demonstrated to improve the accuracy of the reconstructed energy \cite{hmax}.

\begin{figure*}
  \includegraphics[trim = 5mm 125mm 100mm 12mm, clip, width=0.48\textwidth]{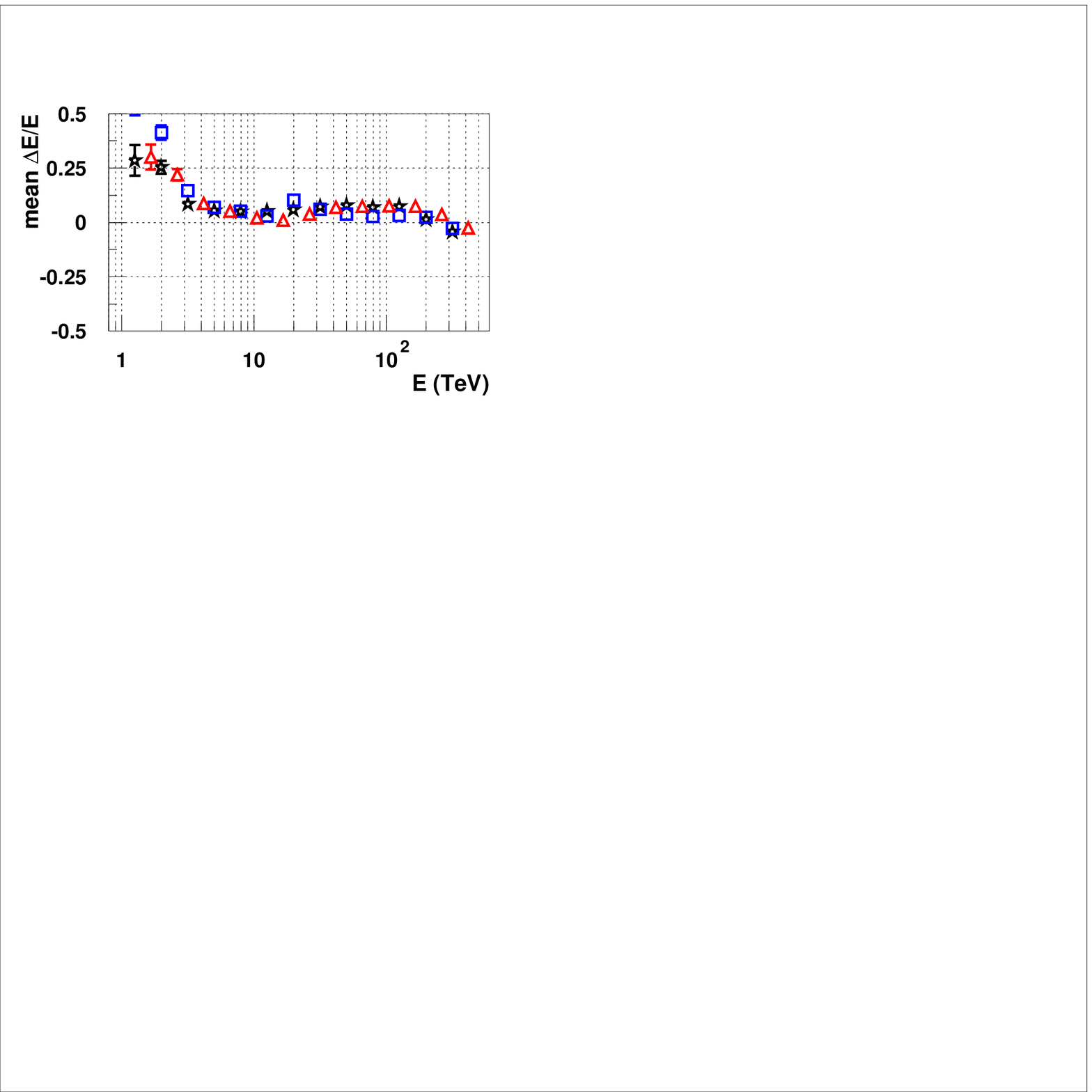}
  \includegraphics[trim = 5mm 125mm 100mm 12mm, clip, width=0.48\textwidth]{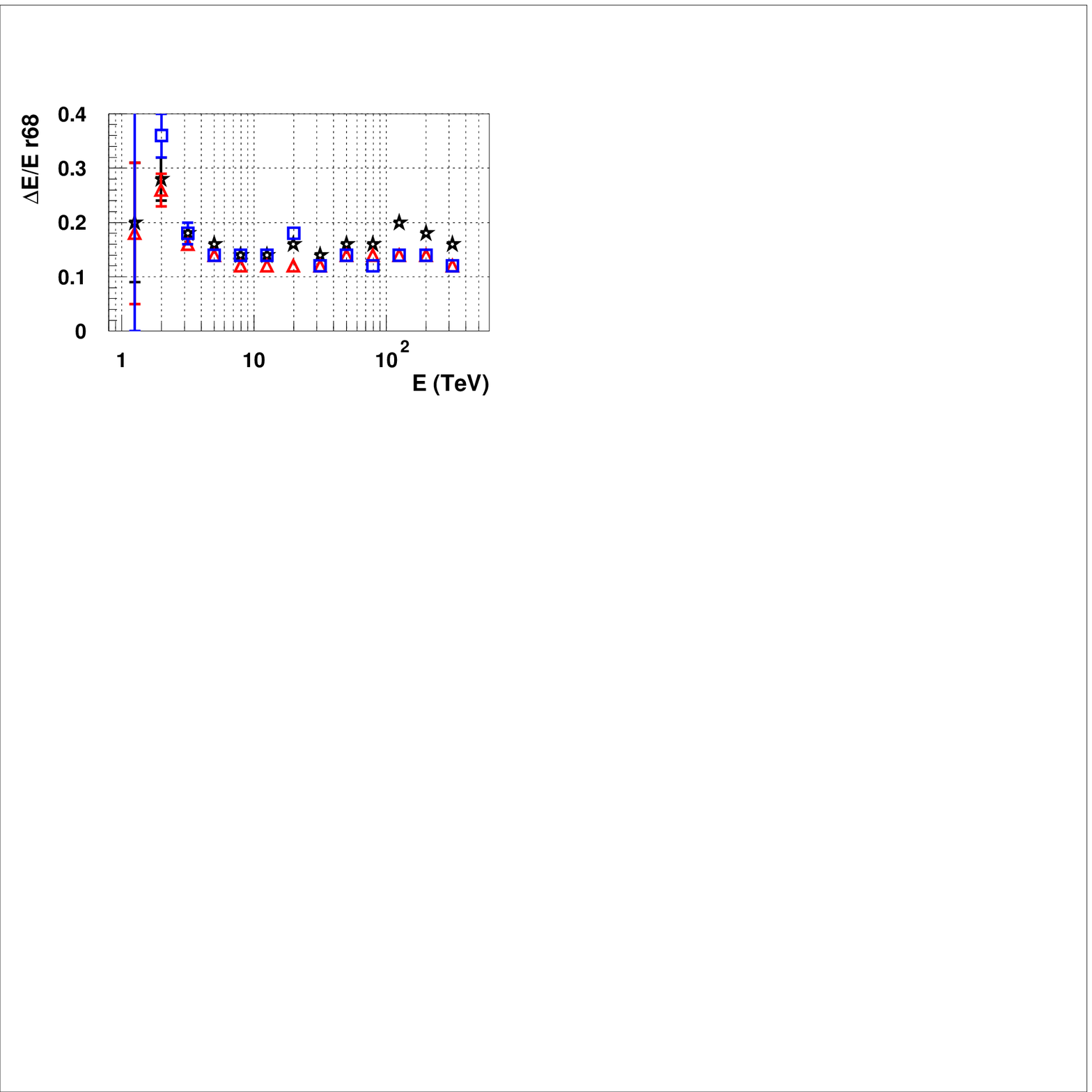}
  \caption{Energy reconstruction bias (left panel) and energy resolution (right panel) of the $(E_{reco}-E_{true})/E_{true}$
    distribution (see text for details), as a function of the true energy, after all shape cuts: MSW$<1.05$, MSL$<1.2$, MSNPix$<1.1$.
    $E_{reco}$ is the reconstructed energy and $E_{true}$ is the true energy.
    Black stars denote the Algorithm 1 event reconstruction,
    blue squares denote the \emph{length}-dependent Algorithm 3 event reconstruction
    and red triangles denote the \emph{time gradient} dependent Algorithm 3 event reconstruction.
    For clarity, in the left panel, the red triangles are shifted higher up in energy by 5\%.
  }
  \label{fig:eres}
\end{figure*}

The energy resolution and the energy reconstruction bias, obtained using the two Algorithm 3 core reconstruction methods, are compared in Figure \ref{fig:eres}.
We define `energy resolution', $r68_{E}$, as the radius that contains 68\% of events in the $(E_{reco}-E_{true})/E_{true}$ distribution at each true energy bin,
where $E_{true}$ is the true energy. The error bars are based on the Poisson error in the number of events within the 68\% containment radius.
Both Algorithm 3 implementations lead to a $\thicksim35\%$ better energy resolution above 100 TeV, compared to the purely geometric core reconstruction.
The gain in core resolution given by \emph{time gradient} dependent Algorithm 3 over the \emph{length}-dependent Algorithm 3
does not appear to improve the overall energy resolution. However, further work on the energy reconstruction algorithm may be 
beneficial in exploiting the timing information.

The term `energy reconstruction bias' in Figure \ref{fig:eres} refers to the mean of the $(E_{reco}-E_{true})/E_{true}$ distribution.
The error bars given in that figure are calculated as $r68_{E} / \sqrt{N}$,
where $N$ is the number of events in the underlying distribution.
Thus $r68_{E} / \sqrt{N}$ is analogous to the `standard error in the mean' in each $E_{true}$ bin.
The prominent positive energy reconstruction bias seen at a few TeV for all reconstruction methods
is caused by threshold effects (e.g. see \cite{hmax}),
which are the result of upward fluctuations in pixel values from showers with energies just below the threshold.
This bias can in principle be corrected, however we do not apply such a correction in this study,
since the effect on our $\gamma$-ray results will be negligible above a threshold of $3-4$ TeV.

\subsection{Off-axis angular resolution}

\begin{figure*}
  \begin{center}
    \includegraphics[width=1.0\textwidth]{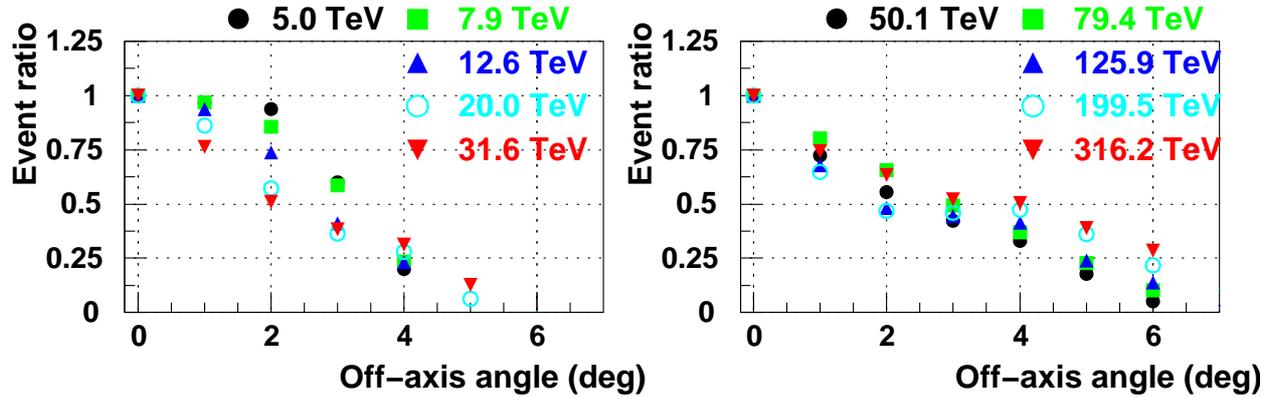}
  \end{center}
  \caption{$\gamma$-ray acceptance as a function of off-axis angle,
    using the \emph{time gradient} dependent Algorithm 3 shower reconstruction.
    The $\gamma$-ray data are divided into logarithmic energy bins whose centres are listed in the two panels above.
    The vertical axis gives the ratio of events that pass shape cuts (MSW$<1.05$, MSL$<1.2$, MSNPix$<1.1$)
    at a given off-axis angle with respect to the number of events passing the same cuts in the on-axis case.}
  \label{fig:acceptances_offaxis_alg3}
\end{figure*}

The large ($4.1^{\circ}$ radius) FoV of the simulated telescopes
allows the detection of showers with a large range of off-axis angles (out to $\thicksim7^{\circ}$).
The energy-dependent off-axis camera acceptance is shown in Figure \ref{fig:acceptances_offaxis_alg3}.
Acceptances are calculated for each off-axis angle as the ratio of off-axis and on-axis post-cut
(MSW$<1.05$, MSL$<1.2$, MSNPix$<1.1$) events,
using the Algorithm 3 core reconstruction in the determination of mean-scaled cut parameters.
Off-axis geometries are obtained by decreasing the telescope axis \emph{altitude} angle
in increments of $1^{\circ}$, starting with $60^{\circ}$, which is the on-axis value.

The off-axis acceptance in Figure \ref{fig:acceptances_offaxis_alg3} has a full width at half maximum (FWHM)
that ranges from $\thicksim 4^{\circ}$ to $\thicksim 8^{\circ}$, depending on the energy.
Given the broadness of the off-axis acceptance, it is important to achieve a good off-axis angular resolution.
Increasing the tilt of the telescopes allows high energy showers with progressively higher core distances to pass event selection cuts.
It was therefore necessary to increase the CORSIKA simulated `throw radius', within which showers are scattered,
from 1 km to 2 km for primary energies above 100 TeV, so as not to artificially limit the number of detected events.
Distant off-axis showers result in a higher proportion of small and/or edge-affected images that are not entirely removed by the quality cuts.
Such images may be poorly parametrized and this, together with lower average telescope multiplicities,
on average, tends to worsen our direction reconstruction.

\begin{figure*}
  \centering
  \includegraphics[height=0.8\textheight]{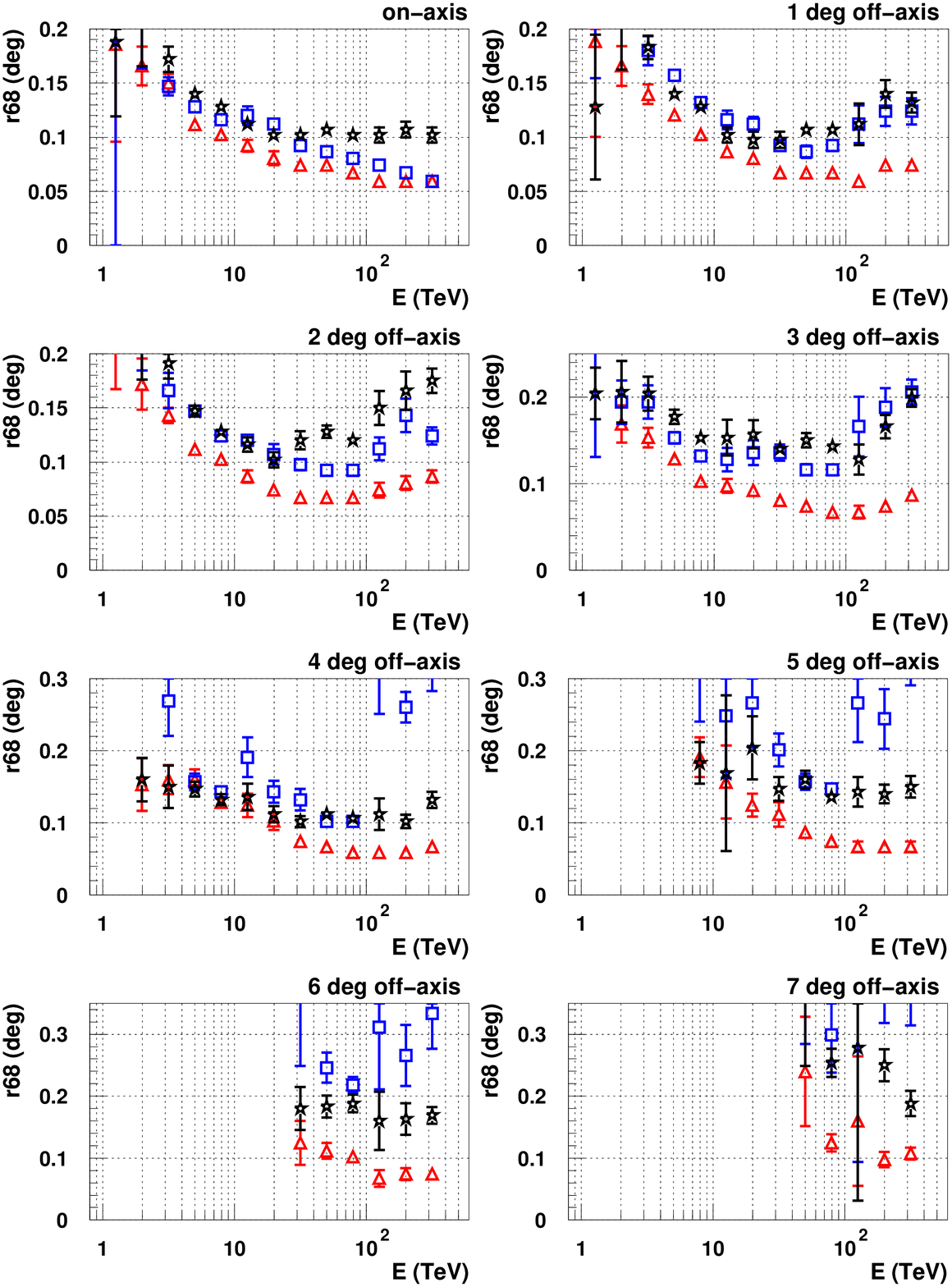}
  \caption{
    Angular resolution $r68$ (radius for 68 \% event containment) as a function of true energy and off-axis angle.
    The $r68$ values are given after all shape cuts (MSW$<1.05$, MSL$<1.2$, MSNPix$<1.1$).
    Algorithm 1, \emph{time gradient} dependent Algorithm 3 and \emph{length}-dependent Algorithm 3 results
    are given as black stars, red triangles and blue squares, respectively.
  }
  \label{fig:offaxis_results}
\end{figure*}

In Figure \ref{fig:offaxis_results}, we compare the off-axis angular resolution achieved using the
\emph{time gradient} dependent Algorithm 3 with that obtained using the \emph{length}-dependent Algorithm 3.
We also show the the results of the Algorithm 1 shower reconstruction.
The energy threshold of the cell rises with increasing off-axis angle,
which leads to reduced event statistics at energies close to threshold (causing larger error bars),
rather than any large change in the inherent angular resolution of the cell.
The general degradation of angular resolution $r68$ above 100 TeV, which is present in all reconstruction algorithms,
is due to the previously mentioned camera edge effects.

Figure \ref{fig:offaxis_results} shows that above $\thicksim$10 TeV,
the \emph{time gradient} dependent Algorithm 3 gives a larger improvement over the \emph{length}-dependent Algorithm 3 than in the on-axis case.
While the \emph{time gradient} is insensitive to increasing off-axis angles,
the behaviour of $d_{t}$ as a function of \emph{length} becomes more degenerate (see \cite{Stamatescu:2010}).
As a result, the distance resolution obtained with \emph{length} and log$_{10}$(\emph{image size})
degrades rapidly beyond $\thicksim2^{\circ}$ off-axis, making this reconstruction method less robust. 
The results of the next-best Algorithm 3 in Figure \ref{fig:offaxis_results} are obtained using distance lookup tables
that are divided into two ranges of $d_{t}$: $<4^{\circ}$ and $>4^{\circ}$. This was chosen `by eye' from the behaviuor of $d_{t}$ as a function of \emph{length}
at $3^{\circ}$ off-axis \cite{Stamatescu:2010}. Adjusting this value (between $\thicksim3^{\circ}$ and $\thicksim5^{\circ}$) for each off-aixs angle
may improve the results, although any differences are expected to be within 10\%.

\section{Conclusion}
\label{section5}

We have presented a shower reconstruction technique for imaging atmospheric Cherenkov telescopes (IACT), which uses pixel timing information.
Our aim was to improve the reconstruction of extensive air showers imaged at large distances ($>200$ m) from the telescopes
of a sparse array. The technique was investigated using a simulated representative array of five modest-sized (6 m diameter),
wide field of view ($8.2^{\circ}$ diameter) telescopes, which would achieve a large $\gamma$-ray collecting area for the multi-TeV energy regime.

The longitudinal air shower time development gives rise to an image \emph{time gradient},
which strongly depends on the distance between a telescope and the position of the shower core on the ground.
This \emph{time gradient} was used to predict the angular distance between the image \emph{cog} and the source position in the camera.
This predicted distance was used in an improved reconstruction algorithm \cite{alg3} (also known as Algorithm 3).

The Algorithm 3 performance was compared to alternative versions that do not use pixel timing information using $\gamma$-ray showers simulated in the energy range of 1-500 TeV.
A noticeable improvement in the angular resolution ($r68$) of the cell was obtained by using the \emph{time gradient} in the reconstruction.
The level of improvement over the `next-best version' of Algorithm 3
varied between $\thicksim$10\% and $\thicksim$40\% for on-axis showers, depending on the energy.

The angular resolution obtained with Algorithm 3 was also compared to that obtained with a purely geometric reconstruction method, known as Algorithm 1 \cite{alg3}.
The improvement with respect to Algorithm 1 ranged between $\thicksim10\%$\ and $\thicksim55\%$ (depedning on the energy) before the introduction of timing information,
and between $\thicksim15\%$\ and $\thicksim75\%$ following its introduction.

For energies in excess of $100$ TeV, an $r68$ angular resolution approaching  $0.05^{\circ}$ was achieved with the \emph{time gradient} dependent Algorithm 3,
while the core resolution was better than 10 m.
Using the same method, an $r68$ angular resolution better than $0.1^{\circ}$ was obtained out to $\thicksim6^{\circ}$ off-axis for energies around 100 TeV.
This approach may improve the survey capability and extended source analysis of future instruments such as \emph{TenTen} \cite{Rowell:2008} and CTA \cite{cta_summary}.

While the level of improvement in array performance does depend on the energy range of interest,
on the telescope parameters, and the analysis method employed, we believe our results highlight the potential
of using timing information in stereoscopic reconstruction in a sparse ($\thicksim500$ m spacing) telescope array.
Thus our study demonstrates the potential of large core distance Cherenkov imaging for multi-TeV astronomy,
with angular and energy resolution performance similar to that of H.E.S.S. in its respective energy regime.

\section{Acknowledgements}
\label{section6}

We thank the referees for valuable comments which improved
the manuscript. This work was also supported by an Australian Research
Council Discovery Project Grant (DP0662810)

\appendix
\section{}
\label{appendix}

The following mathematical expressions relate to toy models
for $\gamma$-ray Cherenkov image time profiles,
which are discussed in Section \ref{section3}.
\small{
\begin{eqnarray} 
 dt(x,z)=\frac{1}{c}\;(\sqrt{x^{2}+z^{2}
 \sec^{2}\gamma+2xz\,\tan\gamma \,\cos\delta} \nonumber \\
 ~~~~~-\frac{z}{\cos\gamma}\,+ \nonumber \\
 ~~~~~\frac{h_0}{z}\sqrt{x^{2}+z^{2}
 sec^{2}\gamma+2xz\,\tan\gamma \,\cos\delta}\eta_0(1-e^{-z/h_0}))\,
\label{eq:timeoff}
\end{eqnarray}

\begin{equation}
  \Omega(x,z)=\arctan\left(\frac{\sqrt{x^{2}+z^{2}
      \tan^{2}\gamma+2xz\,\tan\gamma \,\cos\delta}}{z}\right)\, \label{eq:omegaoff}
\end{equation}
\begin{equation}
  \chi(x,z)=\arctan\left(\frac{\sqrt{z\,\tan\gamma \,\sin\delta}}{x+z\,\tan\gamma \,\cos\delta}\right) \label{eq:chioff}
\end{equation}

\begin{eqnarray} 
 dt(x,z)=\frac{1}{c}(\sqrt{z^{2}+(x+z\,\tan(zen))^{2}}
 -\frac{z}{\cos(zen)}\,+ \nonumber \\
 \frac{h_0}{z}\sqrt{z^{2}+(x+z\,\tan(zen))^{2}}\,\eta_0(1-e^{-z/h_0}))\,
\label{eq:timezen}
\end{eqnarray}

\begin{equation}
  \Omega(x,z)=\arctan(\frac{x\,\cos(zen)}{\frac{z}{\cos(zen)}+x\,\sin(zen)}) \label{eq:omegazen}
\end{equation}
}

\bibliographystyle{model1-num-names}
\bibliography{<your-bib-database>}

\end{document}